\documentstyle[aasms4,raulbib,epsf]{article}
\begin{document}

\title{Galaxy Evolution, Deep Galaxy Counts and the Near-IR Cosmic Infrared 
Background.}

\author{R. Jimenez\altaffilmark{1}, A. Kashlinsky\altaffilmark{2,3,4} }
\affil{$^1$Institute for Astronomy, University of Edinburgh, Royal Observatory 
Edinburgh, Blackford Hill, Edinburgh EH9 3HJ, UK}
\affil{$^2$Raytheon STX, Code 685, NASA Goddard Space Flight Center, Greenbelt, 
MD 20771}
\affil{$^3$NORDITA, Blegdamsvej 17, DK-2100 Copenhagen, Denmark}
\affil{$^4$Theoretical Astrophysics Center, Juliane Maries Vej 30, DK-2100 
Copenhagen, Denmark}

\authoremail{raul@roe.ac.uk, kash@nordita.dk}


\def\plotone#1{\centering \leavevmode
\epsfxsize=\columnwidth \epsfbox{#1}}

\def\wisk#1{\ifmmode{#1}\else{$#1$}\fi}

\def\wm2sr {Wm$^{-2}$sr$^{-1}$ }                
\def\w2m4sr2 {W$^2$m$^{-4}$sr$^{-2}$ }          
\def\nwm2sr {nWm$^{-2}$sr$^{-1}$ }              
\def\lt     {\wisk{<}}
\def\gt     {\wisk{>}}
\def\le     {\wisk{_<\atop^=}}
\def\ge     {\wisk{_>\atop^=}}
\def\lsim   {\wisk{_<\atop^{\sim}}}
\def\gsim   {\wisk{_>\atop^{\sim}}}
\def\kms    {\wisk{{\rm ~km~s^{-1}}}}
\def\Lsun   {\wisk{{\rm L_\odot}}}
\def\Msun   {\wisk{{\rm M_\odot}}}
\def\Zsun   {\wisk{{\rm Z_\odot}}}
\def\um     {$\mu$m}
\def\sig    {\wisk{\sigma}}
\def\etal   {{\sl et~al.\ }}
\def\eg     {{\it e.g.\ }}
\def\ie     {{\it i.e.\ }}
\def\bsl    {\wisk{\backslash}}
\def\by     {\wisk{\times}}

\def\amin   {\wisk{^\prime\ }}
\def\asec   {\wisk{^{\prime\prime}\ }}
\def\cc     {\wisk{{\rm cm^{-3}\ }}}
\def\deg    {\wisk{^\circ}}
\def\ddeg   {\wisk{{\rlap.}^\circ}}
\def\damin  {\wisk{{\rlap.}^\prime}}
\def\dasec  {\wisk{{\rlap.}^{\prime\prime}}}
\def\approxeq{$\sim \over =$}
\def\abouteq{$\sim \over -$}
\def\percm{cm$^{-1}$}
\def\percmsq{cm$^{-2}$}
\def\percmcub{cm$^{-3}$}
\def\perhz{Hz$^{-1}$}
\def\perpc{$\rm pc^{-1}$}
\def\persec{s$^{-1}$}
\def\peryr{yr$^{-1}$}
\def\te{$\rm T_e$}
\def\tenup#1{10$^{#1}$}
\def\to{\wisk{\rightarrow}}
\def\thin{\thinspace}
\def\uk{$\rm \mu K$}
\def\p{\vskip 13pt}

\begin{abstract}
Accurate synthetic models of stellar populations are
constructed and used in evolutionary models of stellar populations in forming
galaxies.  Following their formation, the late type galaxies are assumed to
follow the Schmidt law for star formation, while early type galaxies are
normalized to the present-day fundamental plane relations assumed to mimic the
metallicity variations along their luminosity sequence. The stars in disks of
galaxies are distributed with the Scalo IMF and in spheroids with the Salpeter
IMF. We show that these assumptions reproduce extremely well the recent
observations for the evolution of the rate of star formation with redshift. We
then compute predictions of these models for the observational data at early
epochs for various cosmological parameters $\Omega, \Omega_\Lambda$ and
$H_0$. We find good match to the metallicity data from the damped $L_\alpha$
systems and the evolution of the luminosity density out to $z\simeq 1$.
Likewise, our models provide good fits for low values of $\Omega$ to the deep
number counts of galaxies in all bands where data is available; this
is done without assuming existence of extra populations of galaxies at high
$z$. Our models also match the data on the redshift distribution of galaxy
counts in $B$ and $K$ bands. They also provide good fits to the observed colors.
We compute the predicted mean levels and angular
distribution of the cosmic infrared background produced from the early
evolution of galaxies. The predicted fluxes and fluctuations are still below
the current observational limits, but not by a large factor.  Finally, we find
that the recent detection of the diffuse extragalactic light in the visible
bands requires for our models high redshift of galaxy formation, $z_f
\geq$(3-4); otherwise the produced flux of the extragalactic light at optical
bands exceeds the current observational limits.
\end{abstract}

\newpage
\section{Introduction}

The epoch and process of galaxy formation are still a matter of considerable
debate despite substantial recent observational and theoretical progress. On
the observational side, it is becoming increasingly clear that galaxies must
have formed early on in the evolution of the Universe with the farthest
galaxies known to date having redshifts of 5 and beyond (\pcite{Franx+97};
\pcite{Dey98}; \pcite{Hu98}). Similarly, such high redshift of
formation of the first stellar populations in galaxies is indicated by the
existence of galaxies at moderately high redshifts, $z \sim 1.5$, but which
contain old stellar populations of about 3.5-4 Gyr (\pcite{Dunlop+96},
\pcite{Dunlop_98}).  On theoretical front, the existence of galaxies at these
redshifts coupled with the data on the present day large-scale galaxy distribution 
allow one to reconstruct the spectrum of the pregalactic density
field (\pcite{Kash_98}) independently of any assumed cosmological paradigms.

Other data do not involve record numbers for redshifts of still only a handful
of galaxies, but are just as important.  Such data come from the deep galaxy
counts probing galaxies in various spectroscopic bands, from blue to
near-infrared. It too constrains both galaxy evolution and global cosmological
parameters. The data on galaxy counts are now coupled with the newly obtained
measurements of the luminosity density produced by galaxies in UV, V and J
(1.25 micron) bands at $z\leq 1$ from \scite{Lilly+96} and the new data on the
evolution of cosmic abundances out to high redshift
(\pcite{Pettini+97}). Furthermore, there are now independent measurements of
the star formation rate out to high redshifts (\pcite{Madau+96}); these show
that star formation increases out to $z \simeq 2$ with a possible peak at $z
\sim$ (2-3).

All such data allow one to reconstruct the early evolution of galaxies and
stellar populations in the Universe thereby providing an important test of
galaxy formation processes and the underlying cosmology.  In computing such
evolution one must necessarily normalize galaxy populations to the present-day
data, e.g. galaxy luminosity function in the relevant bands and the
morphological mixes.  The present-day luminosity function of galaxies has now
been measured accurately in B (\pcite{Loveday+92}) and K bands
(\pcite{Gardner+97}) and the morphological mixes at the present epoch are also
well determined (\pcite{Marzke+94}). With the input of the Initial Mass
Function (IMF) for the various galaxy types and specifying their star
formation history, one can uniquely compute individual galaxy populations out
to very early times.

Individual galaxy observations, however, provide only a limited amount of
information on the overall evolution of galaxies and the Universe, and are
expensive in terms of time involved and area covered. On the other hand,
diffuse background radiation fields left over from galaxy formation and
evolution contain cumulative information about the entire evolution of the
Universe including radiation from objects inaccessible to telescopic studies.
Cosmic Infrared Background (CIB) occupies a unique space among the various
diffuse background produced by galaxies. The reason is that the bulk of any
stellar light emitted at early times will reach the observer shifted into the
near to far-IR. CIB thus probes early galaxy formation and evolution and
contains cumulative information on the history of the Universe over redshifts
between the epoch of the last scattering surface, probed by the microwave
background, and $z\sim 0$, probed by the surveys in visible bands.

On the observational side, no detection of the putative CIB has been made at
wavelengths below 100 $\mu$m. At wavelengths beyond 150 $\mu$m there are
claims of possible detections from the COBE DIRBE (\pcite{Hauser+98};
\pcite{Schelegel+98}) and FIRAS (\pcite{Puget+96}) maps. In the near-IR,
arguments based on chemical evolution predict levels of CIB around $\sim 10$
\nwm2sr (\pcite{Stecker+77}). The Galactic and zodiacal foregrounds are very
bright at these wavelengths so it is very difficult to reach such limits
directly as analysis of d.c. levels of DIRBE maps indicates
(\pcite{Hauser+98}). On the other hand, comparable levels can be reached with
fluctuations analysis of the near-IR CIB (\pcite{Kash1+96}; \pcite{Kash+96};
\pcite{KMO98}) and there is hope of
reaching the CIB levels directly with this method applied to other surveys. In
optical bands, \scite{Vogeley_98} has derived strong limits of the diffuse
background at $R$ and $B$ from the fluctuations analysis of the Hubble Deep
Field.

At the same time, significant progress has been made in the last few years in
theoretical understanding of evolution of stellar populations as new opacities
have become available (OPAL 95 and Alexander, 1998 private communication)
for computing stellar interiors from the contracting Hayashi track to the
white dwarf (or carbon ignition) phase. Photospheric modeling has reached high
accuracy with theoretical models resembling very closely observed spectra from
individual stars (\pcite{Jimenez+98}). Furthermore, the new Hipparcos data
allow for a much better calibration of the isochrones at the main-sequence
(see \pcite{Jimenez_Hipp+97} for more details) in conjunction with the
classical calibration to the Sun. This then allows to construct
self-consistent stellar population models that are properly calibrated to the
Sun and to the Hipparcos data and therefore are more reliable than previous
models. Also new stellar yields are now available (see section 4.1) and
chemical evolution can be done more accurately.

It is therefore, imperative and timely to explore the limits and predictions
that models of stellar evolution make in light of the new data, such as
chemical contents at early epochs, the evolution of star formation and
luminosity density with time, as well as the spatial and other properties of
the near-IR CIB and optical diffuse backgrounds.

The outline of this paper is as follows: in Section 2 we discuss the
cosmological context for the calculations we present in this paper. Section 3
discusses the stellar populations models we use for galaxy evolution. Modeling
evolution of different galaxy types which at the same time is normalized to
the data on modern galaxies and the redshift of galaxy formation is presented
in Section 4. In Section 5 we compare our results to the data: we reproduce
the evolution of chemical abundances at high $z$, the data on the luminosity
density in $UV, B$ and $J$ bands from \scite{Lilly+96}; the available data on
deep galaxy counts in the various bands; and in in Section 5.4 compute
properties of the near-IR CIB for the models that successfully reproduce the
above data. Our conclusions are summarized in Section 6.

\section{Cosmological context}

Galaxy evolution models are now constrained by the recent data containing
cumulative information on galaxy evolution at early epochs. In this paper we
concentrate on and aim to construct galaxy evolutionary models to constrain
the following: 1) the evolution of the total emissivity in the various bands
with redshift. This data has been recently obtained by \scite{Lilly+96} from
the CFRS faint galaxy survey in three bands (UV, B and J) out to $z\leq 1$. 2)
The data on deep galaxy counts in different photometric bands.  3) The current
limits on the DC levels and fluctuations in the CIB and diffuse background
light at optical wavelengths. While the current data on the near-IR CIB has
not yet resulted in firm detections, the latest upper limits come very close
to the cosmologically interesting levels. On the other hand, in optical bands
there is now positive detection of the diffuse background light
(\pcite{Bernstein98}). We therefore
aim to accurately compute the expected levels of the near-IR CIB and its
structure that can be uncovered in future searches.

We define with $L_\lambda(\lambda;z)$ the luminosity per wavelength interval
$d\lambda$ centered on wavelength $\lambda$ in the \underline{rest-frame of
the galaxy} located at redshift $z$. The emissivity per unit wavelength
produced by galaxies at redshift $z$ and observed today through a filter with
band-width $d\lambda$ centered at wavelength $\lambda$ is:
\begin{equation}
\epsilon_\lambda(z) = (1+z)^{-1} \sum_i \int
L_{\lambda,i}(\frac{\lambda}{1+z};z) dN
\end{equation}
where $dN$ is the comoving number density of such galaxies which is given by
the luminosity function of galaxies at $z$. If the number of galaxies is fixed
since their formation then the latter is given by the present-day blue (or
red) luminosity of galaxies, i.e.  $dN=\Phi_0(L_B) dL_B$. In our calculations
we compute the emissivity after normalizing to the data on the present-day
galaxy luminosity function in \underline{both} blue ($B$) and red ($K$)
bands. The sum in eq. (1) is taken over all galaxy types. The emissivity per
unit frequency is $\nu\epsilon_\nu(z)=(1+z)^2 \lambda\epsilon_\lambda(z)$.

Another set of quantities that depend on galaxy evolutionary history in a
variety of bands, but in a different way than the emissivity, are deep counts
of galaxies.  The data is now available in four bands: $B,R,I,K$ (e.g. see
\pcite{Koo_Kron_92}, \pcite{Ellis97} for reviews).  
The number of galaxies, ${\cal L}$, per unit solid angle
observed through the aperture in the rest frame band $j$ in the apparent
magnitude interval $[m; m+dm]$ is given per $dz$ by:
\begin{equation}
\frac{d {\cal N}_j}{dz} = \frac{dV}{dz} dN(m_j;m_j+dm_j)
\end{equation}
where the comoving volume occupied by unit solid angle in the redshift
interval $dz$ is $dV/dz = (1+z) x^2(z) cdt/dz$. The cosmic time-redshift
relation depends on the global cosmological parameters via:
\begin{equation}
H_0\frac{dt}{dz} =\frac{1}{(1+z)^2\sqrt{1+\Omega z +
\Omega_\Lambda[(1+z)^{-2}-1]}}
\end{equation}

The last two quantities, which reflect still another dependence on cosmic
epoch and evolution, relate to the levels of the CIB produced by such evolving
galaxy populations: the DC (mean) levels of the CIB and the levels of
fluctuations in its angular distribution. The flux received per unit
wavelength in band $\lambda$ from each galaxy with luminosity $L_\lambda$ at
redshift $z$ is $\frac{L_\lambda(\lambda/(1+z);z)}{4\pi
x^2(1+z)^3}$. Multiplying this with $dN$, then integrating over all the
luminosities, summing over all the galaxy types and multiplying by $dV/dz$
gives the flux produced per redshift interval $dz$:
\begin{equation}
\frac{dF}{dz}  = \frac{R_H}{4\pi}
\frac{1}{(1+z)^2} \frac{d(H_0t)}{dz}
[\lambda \epsilon_\lambda(z)] = 
\frac{R_H}{4\pi}\frac{d(H_0t)}{dz}[\nu \epsilon_\nu(z)]
\end{equation}
where $F=\lambda I_\lambda = \nu I_\nu$ is the total CIB flux in band
$\lambda$ received from evolving galaxies.

An additional measure of the CIB produced by galaxies and thereby of the early
evolution of stellar populations is its spatial structure. Until recently, no
attempts have been made to measure CIB fluctuations although some theoretical
calculations existed in the mid to far-IR (\pcite{Wang91}). Recently,
observational progress has been made from studying COBE DIRBE all-sky maps
(\pcite{Kash1+96}; \pcite{Kash+96}) resulting in interesting upper limits, but
so far no firm detections.  Since galaxies producing the CIB are clustered,
their clustering pattern should in turn be reflected in the structure of the
CIB. A convenient way to characterize the CIB fluctuations, $\delta
F({\mbox{\boldmath$\theta$}}) \equiv F({\mbox{\boldmath$\theta$}}) - \langle
F\rangle$, is via its correlation function, $C(\theta) \equiv \langle \delta
F({\mbox{\boldmath$x$}}+{\mbox{\boldmath$\theta$}})\delta
F({\mbox{\boldmath$x$}}) \rangle$. Its two-dimensional Fourier transform is
the power spectrum $P(q) = \langle |\delta F_q|^2 \rangle$, where $\delta
F(\mbox{\boldmath$\theta$})= (2\pi)^{-2} \int \delta
F_q\exp(-i\mbox{\boldmath$q$}\cdot \mbox{\boldmath$\theta$})
d^2\mbox{\boldmath$q$}$.  The rms fluctuation in the the CIB flux on scale
$\theta\simeq \pi/q$ is $\delta F_{\rm rms} \simeq \sqrt{q^2P_2(q)/2\pi}$.

The two-dimensional angular correlation of the CIB is related to the
underlying three-dimensional galaxy correlation function and the rate of flux
production, $dF/dz$, via the projection equation otherwise known as the Limber
equation (\pcite{Peebles_80}).  One can show via the Limber equation that the
angular power spectrum of the CIB, $P_2(q)$, is related to that of the
present-day galaxy clustering, $P_3(k)$, via (\pcite{KMO98}):
\begin{equation}
\frac{d}{dz}[q^2P_2(q)]= 
\left(\frac{dF}{dz}\right)^2 \frac{\Psi^2(z)}{H_0\frac{dt}{dz}}
\Delta^2\left(\frac{q(1+z)}{x(z)}\right)
\end{equation}
where $\Delta(k) = \sqrt{R_H^{-1} k^2 P_3(k)}$, $P_3(k)$ is the
three-dimensional power spectrum of the present-day galaxy distribution, or
the three-dimensional Fourier transform of the two-point galaxy correlation
function, and $\Psi(z)$ is the growth factor accounting for the evolution
(growth) of the clustering pattern with time.  The latter can be scale
dependent since non-linear and linear scales may evolve differently.  The
quantity $\Delta(k)$ is essentially an order-of-magnitude fluctuation in
galaxies over the line-of-sight cylinder of length $R_H$ and diameter
$k^{-1}$. Thus the net fluctuations in the CIB reflect a different dependence
on the rate of CIB production than the mean (DC) flux in eq.(4), and are also
weighed in the integrand with the quantity $\Delta^2$ containing the galaxy
clustering power spectrum.  (The correlation function, or power spectrum,
being the two-point process, depends on the integral over the second power of
$dF/dz$).  Therefore measuring CIB fluctuations in conjunction with the DC
levels will provide important information on the $dF/dz$ dependence in
addition to the total flux levels.

The set of the quantities given by equations (1)-(5) in this Section produces
numbers that have very different dependences (integral and differential) on
the early history of the Universe. Therefore, matching all the quantities to
the data can strongly constrain the evolutionary models for the early
Universe.

\section{Constructing stellar population models}

Tinsley (1980) pioneered the use of models of synthetic stellar populations
in studying early galactic evolution. These were later refined and developed
by \scite{Bruzual_83}. Since then significant progress has been made in the field 
with many theoretical libraries becoming available
(e.g. \pcite{Yoshii_Takahara_88,Bruzual_Charlot_93,Worthey_94})

In order to compute the spectral evolution of galactic stellar populations we
used a new set of synthetic stellar population models that are an updated
version of the previous models constructed by \scite{Jimenez+98}.  The new
models are based on the extensive set of stellar isochrones computed by
\scite{Jimenez+98} and the set of stellar photospheric models computed by
\scite{Kurucz+92} and \scite{Jimenez+98}. The interior models were computed
using JMSTAR15 (\pcite{Jimenez+98}) which uses the latest OPAL95 radiative
opacities for temperatures larger than 6000 K, and Alexander's opacities
(private communication) for $T$ below 6000 K. For stellar photospheres with
temperatures below $8000 K$ we used a set of models computed with an
updated version of the MARCS code (U. J{\o}rgensen, private communication). We
included in these models all the relevant molecules that contribute to the
opacity in the photosphere. Stellar tracks were computed self-consistently,
i.e. the corresponding photospheric models were used as boundary conditions
for the interior models.  This procedure has the advantage that the stellar
spectra are known at any point along the isochrone and thus the interior of
the star is computed more accurately than if a grey photosphere were
used. Therefore we overcome the problem of using first a set of interior
models computed with boundary conditions defined by a grey atmosphere and then
a separate set of stellar atmospheres, either observed or theoretical, that is
assigned to the interior isochrone a posteriori. The problem is most severe if
observed spectra are used because metallicity, effective temperature and
gravity are not accurately known and therefore the position assigned for the
observed spectra in the interior isochrone may be completely wrong (in some
cases the error is larger than 1000 K).  A more comprehensive discussion and a
detailed description of the code can be found in \scite{Jimenez+98}.

An important ingredient in our synthetic stellar population models is the
novel treatment of all post-main evolutionary stages that incorporates a
realistic distribution of mass loss (see \pcite{Jimenez+98} for a more detailed
discussion). Thus the horizontal branch is an extended branch and not a red
clump like in most other stellar population models. Also the evolution along
the asymptotic giant branch is done in a way such that the formation of carbon
star is properly predicted as is the termination of the thermal pulsating
phase.

Using the above stellar input we compute synthetic stellar population models
that are consistent with \underline {chemical evolution}.  The procedure to
build stellar populations consistent with chemical evolution was as follows:

We start by constructing simple synthetic stellar populations (SSP).  We define
an SSP to be a population with homogeneous metallicity and with no further star
formation activity, i.e. all the available gas was exhausted in the first star
formation episode. The initial star formation episode has a duration no longer
than 1\% of the age of the SSP. The procedure to build an SSP is the following:
\begin{enumerate}

\item A set of $10^6$ to $10^{12}$ (depending on the mass of the population in
the model) stellar tracks from our library is distributed according to an
initial mass function (IMF).
\item Next, we find the points in the plane of $T_{\rm eff}$ vs. Luminosity 
where the previous tracks have the required age for the SSP that we want to
model. We then produce a synthetic color-magnitude diagram of the SSP.
\item To each point of this synthetic color-magnitude diagram we assign the
corresponding self-consistent photospheric model. We finally add all the
individual spectra to produce the integrated spectrum of the SSP.
\end{enumerate}

The main advantage of this approach is that it is free of under-sampling
problems in the fastest stages of stellar evolution and that it allows the
construction of realistic synthetic color-magnitude diagrams since the
procedure simply mimics the nature.
SSPs are the building blocks of any arbitrarily complicated population since
the latter can be computed as a sum of SSPs with variable IMFs and
metallicities.  Thus, the total galactic luminosity can be computed as:
\begin{equation}
L_{\lambda}(t)=\int_{0}^{t} \int_{0}^{M_{f}} \int_{Z_i}^{Z_f} 
SFR(Z,M,t_1)\, l_{\lambda}(Z,M,t-t_1)\, dZ\, dM\, dt_1
\end{equation}
where $l_{\lambda}(Z,M,t')$ is the luminosity of a star of mass $M$,
metallicity $Z$ and age $t'$, $t$ is the age of the population to be modeled,
$Z_i$ and $Z_f$ are the initial and final metallicities respectively of the
population, $M_f$ is the largest stellar mass in the population and $SFR$ is
the star formation rate that depends on the mass, metallicity and age of the
population. In order to compute the SFR it is necessary to know the mass
function as a function of time, the evolution of the metallicity in time ({\it
which we properly take into account using our chemical evolution models}) and
the amount of gas left in the galaxy to form stars that is also properly
accounted for by using the chemical evolution models.  For most of the cases
it is possible to split the $SFR$ into two factors:
\begin{equation}
SFR(Z,M,t)=\Phi(M,t) \times g(Z,t)
\end{equation}
where $\Phi$ is the mass function and $g$ is the fraction of gas available at
any time at the appropriate metallicity consistent with chemical evolution to
form stars. Since an SSP is:
\begin{equation}
SSP_{\lambda}(Z,t-t_1)= \int_{0}^{M_f}{\Phi (M,t) \times l_{\lambda} (Z,
M,t-t_1) dM}
\end{equation}
then we obtain:
\begin{equation}
L_{\lambda}(t)=\int_{0}^{t} \int_{Z_i}^{Z_f}
g(t_1, Z)\, SSP_{\lambda}(Z,t-t_1)\,dZ\, dt_1
\end{equation} 

The approach is fairly straightforward and allows the construction of a large
atlas of synthetic stellar population models. Using this approach we have
computed an atlas of synthetic stellar population models for ages between $1
\times 10^6$ to $1.5 \times 10^{10}$ years with a range in metallicities from
$Z=0.0001$ to $Z=0.1$, i.e. from $0.01$\Zsun$\,$ to $5 Z_{\odot}$. Our models
have been extensively used to study related problems (e.g. \pcite{Dunlop+96},
\pcite{Spinrad+97}), and a detailed comparison with other SSP models in the
literature can be found there.

\section{Modeling galaxy evolution}

\subsection{Individual galaxy evolution}

In order to compute the evolution of an individual galaxy, we constructed
an atlas of synthetic population models in the following way:

\begin{enumerate}

\item 
5 different morphological types are used: E/S0, Sab, Sbc, Scd, Sdm/Irr.  For
Sab, Sbc and Scd we separate the contribution from the bulge and the disc.
The bulge is always modeled as a high metallicity object with a fixed
metallicity of 1.5 $Z_{\odot}$ that was formed at the redshift of galaxy
formation (see below) in a star burst of duration 0.5 Gyr. Early type galaxies
are modeled so that they reproduce the fundamental plane relations.  As is
well known star burst models produce evolution for spheroids and ellipticals
which is too red because the UV excess (upturn shortwards of 2000\AA) observed
in these systems (\pcite{Bertola+82}) is not present (\pcite{Dunlop_89}). In
order to compensate for this deficit/excess we introduced a population of very
blue (horizontal branch like) stars that reproduces the present day UV excess
in spheroids and ellipticals (see Fig.~1). We note that the origin of the UV
excess is still a question of open debate, and hence 
we simply adopt the previous empirical approach in
order to reproduce the observed UV excess of today's spheroids and
ellipticals.

\item Each morphological type is modeled in accordance with chemical
evolution.  The evolution of the global metallicity is followed by taking into
account detailed nucleosynthesis prescriptions by including the contributions
to galactic chemical enrichment by stars of all masses (see
\pcite{Matteucci+89}).  The ellipticals are all formed in a very short star
burst with a duration of 0.5 Gyr, this short time-scale is required in order
to reproduce the observed tight color-magnitude and Mg-Fe relation
(\pcite{Matteucci_Greggio_86}) observed in present-day ellipticals and the
wind model is used for their chemical evolution (\pcite{Matteucci_Greggio_86};
\pcite{Arimoto_Yoshii_87}; \pcite{Arimoto+92}); i.e. star formation stops when
the remaining gas is blown away by the high supernovae activity. This model is
successful in reproducing today's abundances of ellipticals.  The population
of ellipticals is formed with a metallicity spread that reproduces the
fundamental plane properties (see next section). Discs are assumed to form by
primordial gas that is accreted around the bulge. We compute the initial surface
density of the discs assuming they are well described by isothermal spheres
(\pcite{Fall+80}; \pcite{Kash_82}) with spin parameters for the dark halo
$\lambda_{\rm spin}=0.03$, $0.05$ and $0.07$ for Sab, Sbc and Scd
respectively. We take the typical half-radius of the model to be
representative of the galaxy since we are modeling galaxies as point sources
and are not interested in the radial dependence.  We assume that the infall
rate is higher in the center than in the outermost regions of the disc and is
proportional to $\exp(-t/\tau)$, where $t$ is the age of the population and
$\tau$ is the typical time for the formation of the disc. We adopted $\tau=2$,
4, 7 and $\infty$ Gyr for Sab, Sbc, Scd and Sdm/Irr respectively. The star
formation rate is assumed to depend on both the surface gas density, $\Sigma_g
(t)$, and the total surface mass density of the luminous material, $\Sigma_m
(t)$.  In particular, we adopted a law of the type $g(t)= \nu [\Sigma_g
(t)]^{k_1} [\Sigma_m(t)]^{k_2}$ with $k_1=1.5, k_2=0.5$ and $\nu=1$ Gyr$^{-1}$;
the latter value for $\nu$ is found to be the best for the Galaxy
(\pcite{Chiappini+97})). This is essentially the Schmidt law for star
formation in the present-day disks. We further assume that the star formation
law is universal. 

\item We adopted a Salpeter IMF ($x_s=1.35$) for bulges and E/S0 and a Scalo IMF
(\pcite{Scalo+86}) for the discs of Sab, Sbc and Scd as well as for Sdm/Irr. For
disk galaxies the proportions of the stars with each IMF were taken to be like
those of bulge to disk masses. We adopted the following bulge-to-disc mass 
ratios: 0.8, 0.5 and 0.1 for Sab, Sbc and Scd respectively. 

\item No absorption by dust has been included in the modeling.

\end{enumerate}

Fig.1 shows the spectrum for the different galaxy types according to the
adopted prescription. E/S0 galaxy spectrum is shown for 1.5\Zsun$\,$ which as
discussed in the next section corresponds to the metallicity of an $L_*$
galaxy. The dotted lines show the spectrum 1 Gyr after the first burst of star
formation and solid lines show the spectrum 14 Gyr after the first star burst.
Fig.~1 shows that the strongest effect in the evolution of the spectra for the
different morphological types occurs in the region below 4000 \AA. The reason
for this is that the position of the main sequence turn off of the population
is the most sensitive point to changes in age and metallicity (i.e. different
star formation schemes) while the giant branch position remains basically the
same. 90\% of the light in the region below 4000 \AA$\,$ comes from stars near
the main sequence turn-off and below, while beyond 5000 \AA$\,$ the population
is dominated by stars from the giant branch and the dwarfs. For Sdm/Irr there
is very little evolution with time since they have continuous star formation
and therefore there is a continuous reservoir of gas to form young and bright
stars. On the contrary, for E/S0 the time evolution is the strongest among all
morphological types since their gas reservoir is exhausted after the first
star burst. Therefore the population in early type galaxies evolves like
Luminosity $\propto$ age$^{-3}$ on the main sequence, i.e. in the region below
4000 \AA. Indeed at later epochs (age $>$ 10 Gyr) the horizontal branch will
make a significant contribution since it will become extended into the blue
bands.

\subsection{Normalizing early-type galaxies to the fundamental plane}

The previous section described the luminosity evolution of individual galaxies
for the various galaxy types. The luminosity history depends on both the
galaxy age and its metallicity; the dependence is stronger in the visible
bands.  In principle, for late type galaxies for which we model chemical
evolution, the galaxy luminosity at the relevant epochs is practically
independent of the initial metallicity. But for early type type galaxies
(E/S0), which are assumed to have undergone only the initial burst of star
formation and to have remained at the value of $Z$ produced in the brief
initial burst of star formation (e.g. until the wind from supernovae exhausted
the available gas for forming stars) the dependence on the metallicity is more
important. Indeed, as the (initial) metallicity changes between 0.01\Zsun$\,$
and 2\Zsun, the $B$-band luminosity changes by almost 2 magnitudes by about 12
Gyr after the burst. In the $K$-band the dependence is weaker, but must
nevertheless be corrected for.

Therefore, we have to normalize the early type galaxies to a particular value
(or set of values) of $Z$. The simplest assumption would be to choose a
particular and constant value of $Z$ for all early type galaxies. However,
this would not be in agreement with observations indicating substantial
variations in metallicity among the early type galaxies. Furthermore, with
this assumption it would be difficult to reproduce the fundamental plane
relations between the central velocity dispersion ($\sigma$) and the blue
photometric radius of early type galaxies (\pcite{Dressler+87};
\pcite{Djorgovski_Davis_87}).

The $B$-band fundamental plane relations imply that the blue band
mass-to-light ratios of early type galaxies scale with the blue luminosity as
$M/L_B \propto L_B^\kappa$ with $\kappa \simeq 0.25$
(\pcite{Jorgensen+96}). The mass here refers to the total mass inside the
photometric radius where it is dominated by stellar populations.  One could be
tempted to account for this dependence assuming different ages for stellar
populations in the different luminosity (or mass) early type
galaxies. However, since at late times ($t\gg \tau$) the blue luminosity
$L_B\propto t^{-1}$, reproducing the fundamental plane relations in this way
would require
gaps of $\sim$ 10 Gyr between formation of (or the first star burst in) the
low-luminosity ellipticals and the bright ones.

A plausible alternative assumption, which will also allow us to normalize the
metallicity for the early type galaxies, is to assume that the fundamental
plane relations reflect the difference in metallicities, rather than age,
along the luminosity sequence of ellipticals. The differences in metallicity
would imply systematic variations in the IMF as function of galaxy's
mass. Since, the bulk of metals produced in the initial burst come from
massive stars, the necessary variations arise only from the massive end of the
stellar mass function. This hypothesis is not new and was widely discussed
elsewhere (\pcite{Renzini_Ciotti_93}; \pcite{Worthey+96}; \pcite{Franx++97}). In
order to reproduce the fundamental plane relations within the framework of our
models and thereby normalize the early-type galaxies to observational data on
the fundamental plane we need to be able to reproduce (within the framework of
our models) also the recent data on the fundamental plane in $K$
(\pcite{Pahre+95}) which indicate that $M/L_K \propto L_K^{0.16}$. Therefore,
the metallicity variations along the mass sequence of the early type galaxies
should reproduce $L_B \propto L_K^b$ with $b=1.16/1.24\simeq 0.9$. In
addition, the recent measurements of the fundamental plane evolution out to
$z\sim 0.5$ (or $\simeq$ 1/3 look-back-time) indicate that the logarithmic
slope of the dependence remains the same while the amplitude (or mass-to-light
ratio) changes (\pcite{Kelson+97}).

The left panel of Fig.2 shows the dependence of the blue-luminosity on the
metallicity for the models considered in this paper. The lines are drawn for
three values of the time elapsed since the star burst: 9, 12 and 14 Gyr. The
middle panel shows the dependence of the $K$-band luminosity on $Z$ and the
right panel shows the corresponding $L_B$ vs $L_K$ relation for the same
metallicity galaxies. The net result is that it is possible to account for the
fundamental plane relations within the framework of our models if one
considers early-type galaxies to span the range from about 0.15\Zsun$\,$ to
2.5 \Zsun$\,$ for systems older than 9 Gyr. If the time elapsed since the last
star burst is as short as 9 Gyr this would shift the span of metallicities
from $\simeq$ 0.1\Zsun$\,$ to 1.5\Zsun$\,$. In any case, the covered span is
consistent with the range of metallicities observed in elliptical
galaxies. Furthermore, the increase in the $L_B$ (or $L_K$) due to aging of
stellar populations that we find between 5 and 14 Gyr is consistent with the
measurements of the fundamental plane relations out to $z\sim 0.6$
(\pcite{Kelson+97}; \pcite{Jorgensen+97}).

We therefore normalize our early type galaxies so that they reproduce the
fundamental plane relations by varying their metal content along the mass
sequence. For normalization point we choose an $L_*$ galaxy to have absolute
blue luminosity $M_{B,*}=-19.8+0.5\lg h$ (\pcite{Loveday+92}) and central
velocity dispersion $\sigma_* =225$ km/s (\pcite{Davies+83}). The metallicity
of the $L_*$ galaxy is adopted from the Mg-$\sigma$ relation for ellipticals
(\pcite{Bender+96}) to be 1.5 Z$_{\odot}$.

\subsection{Normalizing to the epoch of galaxy formation}

The previous two sections specify evolution of the individual galaxies of both
early and late types once the IMF of stellar populations is assumed. In our
computations we adopt the IMF for the various galaxy types as discussed in
Section 4.1, but note that the results depend somewhat strongly on the assumed
IMF. Our adopted IMF is motivated by reproducing the chemical abundances of
the Galaxy and ellipticals. According to this we chose the Scalo IMF for the
disc systems that is the one that best reproduces the chemical gradients in
the Galaxy (\pcite{Chiappini+97}); likewise the Salpeter IMF is more
appropriate to reproduce the abundances of ellipticals (\pcite{Bender+96}).

Fig.3 shows the time evolution of the $B,I,K$ luminosities with time since the
first burst of star formation. The three bands were chosen since in the rest
frame of the galaxy they reflect emission by very different stellar
populations. $K$ band is dominated by emission from old stellar populations,
$B$ band is dominated by emission from the young hot stars and the $I$ band
emission reflects stars of the intermediate mass and age. In particular the
$B$ band reflects the current star formation activity in the population since
80\% of its light is dominated by the brightest stars on the main sequence
turn-off; therefore the younger the stars the smaller $M_B$. On the other hand,
$K$ band is purely dominated by stars in the hydrogen shell and double shell
burning stages, i.e. the red giant and asymptotic branch as well as by the
dwarfs. Since the evolution along these stellar stages is weakly dependent on
time, it is expected that $M_K$ evolves very little, in agreement with Fig.~3.

One can see that at late times ($t > \tau$) the blue band luminosity decreases
linearly with time. This was discussed by \scite{Tinsley_80} who found that for
Salpeter IMF with logarithmic slope $x_s$ the blue luminosity at late times is
$L_B \propto t^{1.3 - 0.3x_s}$. This dependence is a consequence of the fact
that for Salpeter IMF the total blue luminosity at any given time is dominated
by the luminosity from the most massive stars still on the main sequence,
while the lifetime of the stars in this mass range is $t\propto
l_{star}^{-1.25}$. On the other hand Sdm+Irr are dominated during their entire
lifetime by very massive and new born stars ($M > 10 M_{\odot}$).  In $K$ band
the luminosity changes very little with time at late epochs. This is to be
expected in the bands that are dominated by stellar populations whose change
in luminosity during the Hubble time is small, i.e. giant branches and dwarfs
on the main sequence.  Since, the $K$ band at the same time probes the
bolometric luminosity, this means that the \underline{total} luminosity
emitted by galaxies in their rest frame changes little at late times.

The right panel on the top of Fig.3 shows the evolution of the
\underline{total} blue mass-to-light ratio of galaxies plotted in solar units
for various galaxy types.  Solid line plots the evolution of the $L_*$ early
type galaxy, which has roughly the solar metallicity.  One can see that the
mass-to-light ratio is the highest for early type galaxies and decreases with
increasing $\tau$. This trend is in agreement with observations of the stellar
mass-to-light ratios in galaxies (e.g. \pcite{Faber_Gallagher_79}). One can use
the plot to constrain both the age of the Universe and the epoch of galaxy
formation (cf. \pcite{Franx++97}).  Indeed, the models indicate that at late
times $M/L_B \simeq A t$(Gyr) where $A=1.6, 1.4, 0.85, 0.57, 0.3$ solar units
for E/S0, Sab, Sbc, Scd, Sdm+Irr types respectively and time is measured in
Gyr.  Observations indicate that the blue mass-to-light ratios for stellar
components of galaxies today are $\simeq 16h, 12h, 9h, 7h, 2h$ for the same
type galaxies (\pcite{Faber_Gallagher_79}). This implies that the age of the
galaxies must be $t({\rm Gyr}) \simeq$(10-12) $h$Gyr old requiring a high (and
same for all type galaxies) redshift of galaxy formation. (It is interesting
to note that the age of galaxies derived from this argument scales as $h$
whereas the age of the Universe scales as $h^{-1}$).  The $K$ band
mass-to-light ratio varies very little with time after $\sim 5$Gyr since the
initial burst of star formation. There is also little variation with the
morphological type.  E.g. at $\sim 10$Gyr the difference between E/S0 and
Scd+Irr galaxies is only $\sim$0.5 mag or about a factor of 1.6 in the
$M/L_K$. This is consistent with observations (cf. Table 2 of
\pcite{Faber_Gallagher_79}). Furthermore, our models give $M/L_K \simeq 2$ for
$L_*$ elliptical galaxies after 10 Gyr consistent with observational value of
$\sim 3h$ (\pcite{Dokkum_Franx_96}).

Fig.4 plots the evolution of the luminosity density at three wavelengths:
0.28, 0.44 and 1 micron for the galaxy evolution specified in this section. In
producing the lines in Fig.4 we further assumed that the number density of
galaxies is given by the present day blue luminosity function taken from
\scite{Loveday+92}. No cosmology was assumed here, i.e.  the figure plots
the emissivity in the rest-frame of the galaxies. Note that the evolution in 
time of the luminosity density is in good agreement with previous work by 
\scite{Fall+96}; \scite{Totani+97}.

Fig.5 shows the cumulative star formation rate for the modeling assumed here
vs $z$ for the open Universe with $\Omega=0.2$ and assuming that all galaxies
started forming their stellar populations at $z_f=5$.  The cumulative rate was
computed by turning $g(t,Z)$ (see section 4.1) into $g(z)$ and using the
luminosity function from \scite{Loveday+92}. Dashed line corresponds to the
star formation from only early type galaxies and solid line from all galaxies.
The data points with the errors bars are measurements of star formation rate
from \scite{Madau_97} using local surveys and data from the Hubble deep field
for $z > 2$. The two points at $z > 2$ have been corrected to account for dust
obscuration (see \pcite{Pettini_dust+97}). The square corresponds to the 
SCUBA/HDF \scite{scuba_98} sub-mm survey. 
Conclusions regarding star formation 
at
high $z$ should be treated carefully since these points may really be only the
lower limits for the real star formation rate (see
e.g. \pcite{Dunlop_sfr98}). The agreement is perfect with the SCUBA/HDF data 
point. 
The agreement is also excellent if one assumes that the
dashed line reflects the galaxy populations for which the \scite{Madau_97}
analysis applies (i.e. predominantly late type galaxies with ongoing star
formation).  On the other hand, if the $z > 2$ points are only the lower
limits, then the solid line would agree as well. Otherwise, the solid line can
be easily adjusted to agree with the \scite{Madau_97} data by simply shifting
the redshift of formation of E/S0 and bulges to $z > 5$. This would result in
a second peak at $z > 5$, which still remains unseen but would be supported by
the recent discovery of $z > 5$ galaxies (\pcite{Dey98}; \pcite{Hu98})
and the lack of other than passive evolution of the fundamental plane
(\pcite{Jorgensen+97}) of E/S0 galaxies.  
It is obvious then that the infall
models can account quite accurately for the high redshift points in the
\scite{Madau_97} diagram. But if the SCUBA/HDF point reflects the real star 
formation rate in the Universe for all the galaxy population, our model 
sucessfully reproduces it.    
Because claims to the contrary have been
appearing recently, we feel that it is worth emphasizing that the evolution of
the star formation rate with epoch is not a proof for any particular type of
hierarchical model and neither is it necessarily related to the spectrum of
the initial density field. Rather, it can and, perhaps, should be taken as an
indication of the rate at which the gas inside already formed galactic halos
is converted into stars. 

In principle, our assuming a fixed value of $z_f$ for
all galaxies is an oversimplification. In hierarchical picture galaxy formation
is an ongoing process with small masses collapsing earlier. Furthermore, for
a given spectrum of the primordial density field there would always be (rare)
objects at any given mass that collapsed earlier owing to statistical spread in the
amplitude of their fluctuations. Accounting for such spread will push some of the
galaxies to higher redshifts of formation and make the agreement with the data 
even better.

\scite{Zepf_97} has recently reported the lack of red colors in the field
population at moderate redshifts ($z \approx 1$) which he argues may be
problematic for E/S0 galaxies forming at $z>3$. The argument is based on the
fact that the \underline{purely star-burst} model would produce colors for
E/S0 that are too red compared to what is observed. As previously
discussed in section 4.1 the purely starbusrt model is \underline{not} a good
description of E/S0 since it lacks the UV excess observed in these
galaxies. E.g. \scite{Dunlop_89} has discussed that if one takes the observed
spectrum of a $z=0$ elliptical and simply K-corrects it back to $z=5$, it will
produce colors that are \underline{bluer} than the purely star-burst
model. Therefore, one expects that E/S0 formed at $z>5$ will follow a track in
the color--$z$ plane that is bluer than what the star-burst model
predicts. This would bring formation redshift of $z>5$ for E/S0 in agreement
with observations. In addition, \scite{Zepf_97} has also shown
that a E/S0 formed at $z>5$ can be brought in agreement with observations if
as little as 5\% of the total mass of the E/S0 contributed to late star
formation episodes ($z < 1$). It should also be noted that an initial 
star formation period of 0.5 to 1 Gyr already makes the E/SO bluer than 
the track followed by the \underline{purely star-burst} model thus making E/SOs 
bluer during their whole evolution than Zepf's limit. Furthermore, since 
the outer radii of E/SOs have sub-solar metallicities, colors of the 
integrated light of E/SOs would be bluer than those considered by Zepf 
for the \underline{purely star-burst} which has solar metallicity. Therefore, 
these limits are not necessarily indicative of the redshift of 
formation of E/SOs, since redshift higher than 3 are allowed when considering 
a realistic model for the formation and evolution of stellar populations of E/SOs.

The above calculations carry no cosmological input, only continuing star
formation in late type galaxies and (passive) evolution of early type ones
with time. In order to put these computations in the cosmological context one
has to specify for how long stellar populations in galaxies have been
evolving, or what the epoch of galaxy formation was. For simplicity we will
assume that all galaxies formed first stellar populations at the same epoch
characterized by the redshift $z_f$.  There is currently enough information to
limit the range of possible values of $z_f$. On theoretical side, the
small-scale power in the matter distribution in the currently popular
cold-dark-matter (CDM) models requires low redshift of galaxy formation: $z_f
\simeq$ (3-4) if $\Omega=1$ (\pcite{Efstathiou_Rees_88}) and even lower in the
low-$\Omega$ CDM models (\pcite{Kash_93}).  On the other hand, observations
indicate that the redshift of galaxy formation is much higher than that,
implying the amount of small scale power in excess of that given by the CDM
models (\pcite{Kash_Jimenez_97}; \pcite{Kash_98}). The highest currently
known redshift of quasars, which are believed to be associated with galaxies,
is 4.92 (\pcite{Schneider+91}). By now, there exist directly observed galaxies
at redshifts higher than that: e.g.  the recent record of $z=4.92$
(\pcite{Franx+97}) that comes from a gravitationally lensed galaxy was
superseded by the recent detection of a galaxy at $z=5.34$ (\pcite{Dey98}) with strong 
$L_{\alpha}$ emission. Likewise, existence
of galaxies at intermediate redshifts $z\simeq 1.5$ such as 53W091
(\pcite{Dunlop+96}) and 53W069 (\pcite{Dunlop_98}) with very old stellar
populations of $\simeq$ 3.5 Gyr rule out Einstein De Sitter Universe and for
open models indicate high redshift of galaxy formation which is uncomfortable
for low-$\Omega$ CDM models (\pcite{Kash_Jimenez_97}).  An independent
argument in \scite{Kash_98} based on the fundamental plane properties of early
type galaxies \underline{and} the data on their halo dynamics from the recent
X-ray observations requires similarly high value of $z_f$. In principle, one
can model formation of galaxies of different masses at different epochs as
required in e.g.  CDM-type models. However, this would introduce little
difference to most of our results and since we are interested in studying the
stelar population part of galaxy evolution, in order to accentuate the latter we concentrate
primarily on the stellar modeling effects.

The age of the Universe provides another constraint on the values of
cosmological constants, $(H_0, \Omega, \Omega_\Lambda)$, and the value of
$z_f$. Using the properties of both the luminosity function and morphology of
the horizontal branch it is possible to determine \underline{independently}
the age of the globular clusters. Thus, \scite{Jimenez+96} and
\scite{Jimenez+Padoan97a} have computed the age for the oldest globular
clusters and found a robust value of 12-14 Gyr. This has been recently
confirmed by the new Hipparcos data that have determined an accurate distance
to some Globular Clusters and thus its age, with current values ranging from
12 to 14 Gyr (\pcite{Reid_97}) and an absolute lower limit of 10 Gyr. In
addition, the Galactic disc contains stars older than 10 Gyr
(\pcite{Jimenez_Hipp+97}). These ages would be difficult to reconcile with the
Einstein De Sitter Universe if the values of the Hubble constant are higher
than 50 km s$^{-1}$ Mpc$^{-1}$.

\section{Results vs observational data}

In modeling galaxy evolution at early times, one usually divides the possible
evolutionary tracks into two: luminosity evolution and spectral evolution
(cf. \pcite{Yoshii_Takahara_88}). A further evolutionary complication is the
possibility of mergers of very early galaxies affecting their stellar
populations (\pcite{Broadhurst+92}). On the other hand, there is a significant
bulk of evidence that the dominant evolution comes only from passive stellar
evolution in early-type galaxies and the effects of ongoing star formation in
late-type ones (e.g. \pcite{Barger+98}; \pcite{Driver+98}) since the global
counts of galaxies and their redshift distribution can be easily modeled
without invoking a population of mergers.  We assume here that once the
galaxies have formed, the dominant evolution comes from aging stellar
populations and the prescribed star formation rate that forms new populations,
since as we show these minimal evolution assumptions provide good fits to all
the available data.

This in turn means that the comoving number density of galaxies $dN$ in the
interval of their \underline{present-day} luminosity $dL$ remains the same
since the epoch of $z_f$. We normalize the luminosity function of galaxies to
the present-day on \underline{both} the $B$-band luminosity function from
\scite{Loveday+92} and the $K$-band luminosity function adopted from
\scite{Gardner+97}.

The morphological mixtures used are the ones from the CfA survey redshift
survey with fractions 0.28, 0.19, 0.32, 0.14 and 0.066 for E/S0, Sab, Sbc, Scd
and Sdm/Irr respectively (\pcite{Marzke+94}). These fractions are assumed to
remain constant in time.

In the calculations in this section we consider four values of the
$z_f=3,5,7,10$ and adjust the values of the cosmological parameters to
reproduce reasonable ages for galaxies and the Universe.  We consider
cosmologies with $(\Omega, \Omega_\Lambda)= (1,0), (0.2,0), (0.2,0.8),
(0.4,0.6)$; for these parameters, the age of the Universe is $t_0 = (6.6, 8.3,
10.7, 8.7) h^{-1}$Gyr respectively.  The values of the Hubble constant
that we considered are $H_0 = 50,65,80$ km/sec/Mpc.

\subsection{Chemical evolution and $Z(z)$}

The first measure of the evolution of stellar populations at early times comes
from the measurements of the cosmic metal abundances at high redshifts. Such
data are now becoming available.

Pettini et al (1997) have observed a sample of 34 Damped Lyman $\alpha$ (DLA)
systems between redshift $z \approx 0.5$ to 4 and measured their $Zn$ and $Cr$
abundances.  Because $Zn$ traces quite closely the iron abundance and is not
depleted by dust as $Cr$, they then translated $Zn$ abundance into an average
metallicity, $\langle Z\rangle$, at the above redshift range. They
have also estimated the amount of dust present in these systems and have
concluded that systems at high redshifts are much less metal and dust abundant
than the Galaxy today (Pettini et al 1997). These measurements provide the
most extensive current sample with which we ought to compare our models.

The nature of the DLA systems is still open to debate, since only a few of
DLAs have been imaged in order to resolve their morphological nature resulting
in normal and low surface brightness galaxies.  Nevertheless, it seems
appropriate to compare the Pettini et al. (1997) data with the predictions
from our model for the \underline{discs} of Sab, Sbc and Scd since these will
be the most likely systems to be responsible for the DLAs. Some caveats should
be noted here: firstly, while very important, the Pettini et al. (1997) data
still have large error bars and only a weak $z$ -- $\langle Z\rangle$
correlation, therefore they may not place very strong constraints on the
global evolution of metallicity in the Universe. Secondly, a galaxy is in
reality an extended object with different metallicities at different
radii. Therefore it is not surprising that Pettini et al (1997) data show a
significant dispersion at every redshift, this may be only due to the fact
that galactic discs at different radii are responsible of different DLAs.

In this work we are modeling galaxies as source points so at every redshift
there is no dispersion. Using the prescriptions for chemical evolution given
in section 4.1 and assuming the galaxy mixtures from CfA, it is possible to
(re)construct an average galaxy at every redshift and compute the average 
metallicity.
The average metallicity is simply computed by tracking the stellar yields
produced by the population at the appropriate time (redshift) and taking into
account the inflow of primordial gas falling into the disc
(e.g. \pcite{Matteucci+89}).  Fig.~6 shows the predicted evolution of the
averaged metallicity with redshift. It is clear that our model predictions are
consistent with the data from Pettini et al. (1997) over the entire redshift
range and for all cosmological models. They also reproduce today's average
metallicity for the Galaxy. A certain element of uncertainty may arise in comparing
our results with the data. Indeed, if the damped Lyman-$\alpha$ systems are disks 
they must have significantly more extended disks than present to match the observed
surface density of absorbers. Our models do not account for extended disks, but
given the size of the current error bars for the data, the treatment we present
is probably adequate. 

\subsection{Evolution of the luminosity density}

We have computed the evolution of the luminosity density as described in
section 2 for the cosmological parameters and formation redshifts given at the
beginning of this section. The luminosity density was computed in 3 bands
$UV=2200$ \AA$\,$, $B=4400$ \AA$\,$ and $J=12000$ \AA.  Fig.~7 shows the
evolution of the luminosity density compared with the CFRS data (\pcite{Lilly+96})
as well as
\scite{Gallego+96} and \scite{Connolly+97}. We have normalized our models to the
present day luminosity function in B (see section 2). Solid lines correspond
to $z_f=3$, dotted lines to $z_f=7$ and dashed lines to $z_f=10$; in all cases
we used $H_0=65$ km s$^{-1}$ Mpc$^{-1}$.

The agreement in most cases is quite good and for zero cosmological constant
the lines always pass inside the error bars in all 3 bands. This is indeed a
success of the modeling since the UV band is very sensitive to the current
star formation while the J band is sensitive to the old stellar and weakly
evolving population. On the other hand for $\Omega=0.2$ and
$\Omega_\Lambda=0.8$ the predicted UV flux lies below the observed data.  The
reason for this is that for this cosmological model, galaxies are older at all redshifts
compared with the open model. In this we differ from the results by
\scite{Totani+97} who find that only flat $\Omega=0.2$ and $\Omega_\Lambda=0.8$
model can account for the observed CFRS data thereby ruling out all other
cosmological models. On the other hand, there is good agreement with
\scite{Fall+96} who find numbers similar to those in Fig.7 for the Einstein-De
Sitter Universe, but using a different approach to compute the emissivity. The
difference between \scite{Fall+96} and our models, on the one hand, and
\scite{Totani+97} on the other can be entirely due to different assumptions
about the IMF and the chemical evolution. This is known to lead to very
different predictions for the evolution of the luminosity density. As an
example we show with thick line in the $(\Omega,\Omega_\Lambda)=(0.2,0)$ panel
of Fig.~7 the evolution of the luminosity density for J band using a Scalo IMF
for the whole population of galaxies. It is clear that the prediction
overshoots the data as previously noted by \scite{Madau_97}. But the important
point to note here is that with certain freedom in the choice of the IMF (and
thus the chemical evolution law) it is possible to bring the $\Omega=0.2$ and
$\Omega=0.8$ model in agreement with the CFRS data.

We also used the K band normalization to compute the
luminosity density and found only about 5\% difference. Therefore both
luminosity functions give a consistent picture and most likely trace the same
populations of the modern day galaxies.

\subsection{Deep galaxy counts}

Recent years have brought wealth of data on galaxy counts down to very faint
magnitudes (see \pcite{Koo_Kron_92}; \pcite{Ellis97} for reviews).  
In $B$ band the counts go
down to $B\simeq 28$ (\pcite{Metcalfe+96}); in $R$ they go to $R\simeq 27$
(\pcite{Smail+95}); in $I$ to $I\simeq 26$ (\pcite{Smail+95}); and the recent
measurements in $K$ using the NICMOS array on Keck telescope extended the
previous measurements (\pcite{Cowie+91}) to $K\simeq 24$
(\pcite{Djorgovski+95}). Further measurements come from the redshift
distribution of galaxies at faint magnitudes in $B$ and $K$ bands
(e.g. \pcite{Songaila+94}, \pcite{Cowie+96}, \pcite{Glazebrook+95}).  Up to now
the general view was that it is very difficult to reconcile $B$ and $K$ counts
within the framework of stellar evolution of one galaxy population. This has
lead to construction of models with extra population of galaxies at high
redshifts (\pcite{Cole+92}, \pcite{Babul_Rees_92}) which by now have either died
or merged into the observed populations.

The four panels in Fig.8 plot the latest compilation of the data on deep
galaxy counts in the four bands, $B,R,I,K$. The lines show theoretical fits
from our models; all the lines are shown for $H_0=50$ km s$^{-1}$
Mpc$^{-1}$. Larger values of the Hubble constant will leave less time for the
evolution of stellar populations, making galaxies brighter (cf. Fig.3).
However, we find that the differences in the predicted count values are small
and we do not plot lines with higher values of $H_0$ in the already condensed
Fig.8. Solid lines correspond to the Einstein De Sitter Universe, $\Omega=1,
\Omega_\Lambda=0$; dotted lines to open Universe with $\Omega=0.2$, and dashed
and dashed-dotted lines correspond to flat Universe with cosmological constant
$\Omega_\Lambda=0.8$ and 0.6 respectively. Einstein De Sitter Universe
predicts deficient numbers for the counts at faint magnitudes in all
bands. This happens for several reasons: the cosmic time allowed for evolution
is the smallest here; the volume at high redshifts is small reflecting both
the Eucledian space and short cosmic time; and finally in the given magnitude
range the galaxies are shifted to lower redshifts in the Einstein De Sitter
case leading to further decrease in the volume and time. Open Universe with
$\Omega=0.2$ provides a good fit to $K$ counts, but passes through the lower
end of error bars of the data at the faintest magnitudes in other bands. A
slightly lower value of $\Omega$ in the open Universe will further improve the
fit to the data.

The numbers in Fig.8 are shown for $z_f=5$. For lower values of $z_f$ all the
curves will lie under the data at all bands, whereas for higher values of
$z_f$ the fits for low $\Omega$ models would become even better. As discussed
earlier in the paper, such high redshifts of galaxy formation are indeed
supported by the recent observations.  The theoretical models shown in Fig.8
were normalized to the present galaxy luminosity function in $B$ taken from
\scite{Loveday+92}. An alternative normalization (particularly in the $K$ counts
calculations) is to the data on the present day galaxy luminosity function in
$K$ determination of which has recently become possible. We also computed
the counts in models normalized to the present-day $K$ band luminosity
function taken from \scite{Gardner+97}. This gave essentially the same numbers
as for the $B$ luminosity function arguing for both the consistency of our
results and the fact that the two bands, $B$ and $K$, likely map the same
galaxy populations.

Indeed a further test on the redshift distribution of faint galaxies comes
from their colors, i.e. from photometric determination their redshifts.
\scite{Djorgovski+95} have obtained I, g and r colors for the faintest
galaxies in their sample (see fig.~4 in \pcite{Djorgovski+95}).  Our predicted
K counts for $23 < K < 24$ extend to redshift 5. At this redshift our models
predict $I-K \approx$ (2-3) in excellent agreement with colors measured by
\scite{Djorgovski+95}; this would imply that galaxies with $23 < K < 24$ are
indeed at $z\simeq$(4-5). On the other hand, if galaxies were not at $z
\approx 5$, then their colors (e.g. I$-$K) should be different than those
predicted from galaxy models. Note, however, that ideally one would require
more than just two bands for accurate photometric determination of
redshifts. A much more comprehensive compilation of colours of galaxies with K
magnitudes as faint as $K=22$ is the one by \scite{Moustakas+96}. We have used
their $I-K$ color distribution to test our models. In Fig. 9 we show the data
from \scite{Moustakas+96} (diamonds) compared with our reddest population of
E/S0s (thick lines) and Sbcs (light lines) for two cosmological models.  Both
galaxies were evolved for a total of $10^{12}M_\odot$ in stars. The early type
galaxy model was chosen to reproduce the present-day metallicity of 2.0\Zsun;
it was normalized to lie on the fundamental plane as discussed in sec.4.2. The
lines are shown for redshift of formation $z_f= 10$ in order to accentuate the
reddening. One can see from Fig. 9 that our models are in very good agreement
with the data and also do not produce extremely red colours. It is worth to
note that the E/S0s plotted in the graph is the reddest galaxy that one could
find in our Universe-model since it is formed at $z=10$ (biggest K correction)
and contains the most metal rich population; in addition it evolves passively
after a short period of star formation (of about 1 Gyr). It therefore
represents a strong upper limit on how red in $I-K$ our galaxies can
be. Fig.~9 clearly shows that our model predictions are in excellent agreement
with the data. The reason for the lack of red colors in our galaxy population
models is the extended star formation period for all galaxies and specially
for E/S0s, but which at the same time is 
short enough to allow E/S0s to produce $\alpha$-elements at
super-solar metallicities.

Further constraints come from the redshift distribution of the counts. Such
information is currently sparse, but could become more abundant in the future.
The top two panels in Fig.10 plot the fits of the $\Omega=0.2$ models to the
data in $K$ and $B$ bands with thick solid lines.  
The data are plotted as histograms and are taken for $17 < K < 18$ from
\scite{Songaila+94} (thin solid line), and 
for $22.5 < B < 24$ from \scite{Glazebrook+95} (solid
lines) and \scite{Cowie+96} (dotted line). As the figure shows, theoretical
lines based on our models for $\Omega=0.2$ Universe provide good fit to the
data. In particular, our theoretical models predict the same redshift for the 
maximum as the observations in both in $K$ and $B$ bands. This especially
true in light of the observational uncertainties  illustrated by the
differences between the \scite{Glazebrook+95} and \scite{Cowie+96}. (The latter
dataset shows a high-redhsift tail in excellent agreement with our theoretical
predictions). Our theoretical
models predict a high-z tail of in the K-band distribution of galaxies, 
although it is quite small (5\% of the total number). 
This is reasonable agreement with the \scite{Songaila+94} data particularly so
since they are only
85\% complete. Future more complete surveys will unveil if there is a tail
of high-redshift objects in the $K$ band.

 The good fits of the models to the available data for \underline{both}
the counts and their redshift distribution are quite encouraging and it is
interesting to analyze the redshift distribution of the faintest counts in the
blue and $K$ bands.  The bottom two panels in Fig.10 show the redshift
distribution of the counts from our theoretical models for the Einstein De
Sitter Universe (dotted line) and open Universe with $\Omega=0.2$. The lines
are drawn for the faintest magnitude range of the present day $K$ counts (left
bottom panel) and $B$ counts. One can see that for the open Universe, which
fits the counts, the faintest $K$ counts should be probing galaxies at
$z\simeq$ (4-5). For the Einstein De Sitter model which fails to account for
the counts, such galaxies would lie at significantly lower redshifts (which in
turn could be one of the reasons for the counts deficit in the Einstein De
Sitter Universe). The faintest $B$ magnitudes currently available probe galaxy
distribution at significantly lower redshifts as the right lower panel of
Fig.10 shows.

Adopting the redshift of galaxy formation $z_f >5$ would
increase the low-$\Omega$ curves for the number counts and would also improve
fits to the data.  We conclude that our models for evolution of galaxies and
synthetic stellar population give good fit to the data with low $\Omega\simeq 0.2$ and with or 
without
the cosmological constant.

\subsection{The near-IR cosmic infrared background from early galaxies}

A further measure of galaxy evolution at early times and a subject of
intensive searches is the cosmic infrared background produced by them. There
exist many calculations of the total levels of the CIB produced by galaxy
evolution (e.g. \pcite{Partridge_Peebles_67}, \pcite{Stecker+77},
\pcite{Bond+86}, \pcite{Franceschini+91}, \pcite{Fall+96}, \pcite{ms98}). The models predict
the levels of the CIB in the near-IR around 5-15 \nwm2sr . Fluctuations in the
CIB have been calculated by \scite{Wang91}, \scite{Kash1+96} and
\scite{Kash+96} and generally give the amplitude of $\sim$ (5-10)\% on the
DIRBE beam scales ($\sim$0.5 deg).

We computed the properties of the CIB in our models from eqs.(4) for the mean
level, and from eq.(5) for fluctuations produced by galaxy clustering.  Fig.11
shows the rate of the flux production, $dF/dz$ in different bands for our
models with $H_0=50$ km/sec/Mpc. The four panels correspond to the
values of $z_f$ written above each box. Solid lines correspond to $J$ band
(1.25 $\mu m$), dotted to $K$ (2.2 $\mu m$) and dashed lines to $L$ band (3.5
$\mu m$). Thick lines of each type correspond to $\Omega=0.2$ and thin lines
to the Einstein De Sitter Universe, $\Omega=1$. The general structure of the
figures is the same for all cosmologies, only numerical values (slightly)
differ: there is an initial peak corresponding to the initial burst of star
formation, where the flux is dominated by the redshifted emission from massive
(O,B) stars. After the first massive stars die the flux rate subsides and then
for most values of $z_f$ it reaches a dip when the flux is dominated by the
galactic spectra redshifted from $\lambda \simeq$ (3000-4000) \AA. As Fig.1
shows there is a sharp drop there for galaxy emission from all types of
galaxies (the 4000 \AA$\,$ break is a blend of several absorption lines); the
largest drop is for the early-type galaxies where this part of the spectrum is
not replenished by photons from the newly-forming massive stars. For low
values of $z_f$ the galaxy rest frame wavelengths of $\lambda < 4000$ \AA$\,$
do not get shifted into the $H,K$ bands which explains the absence of the dip
for the dashed and dotted lines in the $z_f=3$ panel.

For higher values of the Hubble constant, the galaxies are younger and, hence,
brighter at each value of $z$. This would lead to the rise in the amplitude of
the curves in Fig.11, but the overall shape would remain the same. Fig.12 shows
the values of the total CIB flux for low (left panel) and high values of the
Hubble constant. The fluxes in \nwm2sr are shown in eight bands: $B$ (0.44
micron), $R$ (0.8 micron), $I$ (0.9 micron), $J$ (1.25 micron), $H$ (1.65
micron), $K$ (2.2 micron), $L$ (3.5 micron) and $M$ (5 micron). At the longest
wavelengths there is a significant contribution from dust in the nearby
galaxies as well as stellar atmospheres and our flux numbers should be
regarded only as lower limits. The different symbols correspond
to the four ($\Omega, \Omega_\Lambda)$
different cosmological models considered; thick
symbols correspond to $z_f=3$ and thin symbols to the same value of ($\Omega,
\Omega_\Lambda)$ and $z_f=10$. One can see that there is little variation in
the total CIB flux with different cosmologies, but at given $z_f$ and $H_0$
the biggest values of flux come from the low-$\Omega$ models which also fit
the galaxy counts. Since the less time the galaxies had to evolve the brighter
they are at each redshift, the flux values increase with increasing Hubble
constant and/or decreasing $z_f$.

In order to compute the power spectrum of the CIB from eq.(5) we have to input
the power spectrum of galaxy clustering today, $P_3(k)$. On small scales ($<50
h^{-1}$ Mpc) it was taken from the APM survey of galaxies in $b_J$ (Maddox et
al 1990) with the power spectrum adopted from the Baugh \& Efstathiou (1993)
de-projection of the APM data on the 2-point angular galaxy correlation data.
On larger scales it was assumed to go into the Harrison-Zeldovich regime
consistent with the COBE DMR measurements (Smoot et al 1992, Bennett et al
1996). Two regimes of the evolution of galaxy clustering were assumed:
clustering stable in proper coordinates, $\Psi^2(z) \propto (1+z)^{-3}$, and
$\Psi^2(z) \propto (1+z)^{-2}$ which roughly corresponds to either growth of
linear fluctuations in the Einstein de Sitter Universe or clustering pattern
with the galaxy two-point correlation function $\xi\propto r^{-2}$ being
stable in comoving coordinates. These describe two extremes for the evolution
of galaxy clustering (\pcite{Peebles_80}).

The upper panels in Fig.13 plot the order-of-magnitude fluctuation in four
bands ($J,H,K,L$) on angular scale $\sim \pi/q$ in the CIB flux,
$\sqrt{q^2P_2(q)/2\pi}$ in \nwm2sr , vs $q^{-1}$ shown in arcmins. Thin lines
correspond to $H_0=50$ km/sec/Mpc. Solid lines correspond to the Einstein de
Sitter Universe, dotted to the open Universe with $\Omega=0.2$, dashes to the
flat Universe with $(\Omega, \Omega_\Lambda)=(0.2,0.8)$ and dashed-dotted
lines to $(\Omega, \Omega_\Lambda)=(0.4,0.6)$.  Two lines of each type
correspond from bottom to top to clustering pattern stable in proper and
comoving coordinates. Thick solid line corresponds to $H_0=80$ km/sec/Mpc. 
One can see that on scales of about $\pi/q \simeq 0.5$ deg the CIB
fluctuations are about 10\%.  On smaller scales, or large $q$, they increase
with decreasing angular scale as $\propto (q^{-1})^{-0.35}$.  On scales above
a few degrees the slope of the CIB fluctuations changes since the power
spectrum there reflects reflects the initial power spectrum of matter
distribution.

Fig.13 shows the contribution to the fluctuations in the CIB from various
redshifts.  The lower panels show $z (dF/dz)^2 \Delta^2(q(1+z)/x(z))$ which
reflects to within a slowly varying factor $\Psi^2(z)/(H_0dt/dz)$, the rate at
which the fluctuations are generated with $z$.  Solid lines correspond to
$q^{-1}=$ 10 arcmin and dotted lines to $q^{-1}=$ 1 degree.  Thick lines
correspond to $z_f=10$ and thin lines to $z_f=5$. The numbers are plotted for
$H_0=50$ km s$^{-1}$ Mpc$^{-1}$. One can see that the CIB fluctuations contain
information about galaxy evolution at high and cosmologically interesting
redshifts. The range of redshifts contributing to each band fluctuations
increases with increasing wavelength and angular scale. The shape of the
lines, including the peak at high redshifts and the following dip, is
determined by the form of $dF/dz$.

We therefore find that for models normalized to the data on galaxy counts, one
can expect significant CIB fluxes of (10-30) \nwm2sr in the near-IR bands of
$J,H,K$, which is larger than previous estimates. Likewise, there should be
significant and potentially measurable fluctuations in the CIB flux, which on
sub-degree scales, are in excess of 10\% of the mean flux.  The current best
limits from the DIRBE all-sky analysis on the CIB (\pcite{Hauser+98}) and its
fluctuations (\pcite{Kash1+96}, \pcite{Kash+96}) are still above our
estimates, but not by a large margin. Note that larger fluctuations could be
produced if the dip in $dF/dz$ due to the redshifted galaxy emission
originating at $\lambda <$ 4000 \AA$\,$ is filled up. This could be achieved
by e.g. a moderate star-burst (5\%) at early stages during the beginning of
the infall process.

A further constraint on the underlying cosmology and galaxy evolution would
come from the measurements of the diffuse extragalactic background in the
optical bands. Such measurements were provided recently by \scite{Bernstein98}
 for the mean (DC) level analysis and by \scite{Vogeley_98} from the
fluctuations analysis of the Hubble Deep Field. Both give fluxes between 10
and 20 \nwm2sr in $B$ (0.44 micron) and $R$ (0.81 micron) bands; the results from
Bernstein (1998) are plotted as filled circles with error bars. We computed
the predicted flux levels in these bands from our models and these are plotted
in Fig. 12 with the same symbols as the near-IR bands. The predicted levels
are in good agreement with the Bernstein (1998) and
\scite{Vogeley_98} measurements if the redshift of galaxy formation is high;
the numbers for $z_f=3$ models would produce too much flux in these bands,
especially if the Hubble constant is high. It is worth pointing out that
because of the Lyman break at 912 \AA$\,$ in the spectrum of galaxies, the
diffuse extragalactic background in the visible bands does not probe as early
epochs as those reflected in the CIB.

\section{Conclusions}

In this paper we considered constraints from and predictions for the early
Universe that follow from the evolutionary models of stellar populations in
forming galaxies. We constructed accurate synthetic models for the evolution
of stellar populations which are assumed to follow the Schmidt law for star
formation. In modeling galaxy evolution we further account for chemical
evolution. Early type galaxies have been normalized to the fundamental plane
relations assumed to reflect variations in the mean metallicity along the
luminosity sequence. Galaxy mixes are adopted from the CfA catalog. We assumed
the Salpeter IMF for stars in early type galaxies and bulges of disk galaxies
and the Scalo IMF for the disk stellar material. The galaxy numbers were
normalized to the present-day galaxy luminosity function measurements in both
$B$ and $K$ bands; both give consistent results.

This allowed us to compute parameters that characterize the evolution of
stellar populations in the early Universe to be compared with the available
observational data. The computations were made for various cosmological
density parameters $\Omega, \Omega_\Lambda$ and the Hubble constant.

Our main conclusions can be summarized as follows:

1) We computed the evolution of the mean cosmic metallicity with $z$. All
models and cosmological parameters provide good fits to the current data. The
current data have substantial uncertainties; after these get reduced one can
hope to be able to further discriminate between the various models and
cosmologies.

2) The evolution of the mean luminosity density in the $UV, B,J$ bands is well
reproduced in our models, but models with zero cosmological constant are
preferred.  Our models reproduce well the recent HST data on the evolution of
the star formation rate with redshift.

3) Our models give good fits to the available data on the deep galaxy counts
in all bands where the observations are available, $B,R,I,K$, for low $\Omega$
models (both flat and open Universe). The models thus do not require
additional galaxy populations at intermediate and high redshifts in order to
explain simultaneously $B$ and $K$ counts. We further fit well the redshift
distribution of $B$ and $K$ counts in the magnitude range where such data are
available. Our models give good fits to the color-magnitude data
distribution, $I-K$ vs $K$, 
and do not produce extremely red colours even if early type galaxies
form at very high redshifts.

4) We compute in detail the mean CIB flux produced by this evolution and the
power spectrum of the CIB angular distribution. Our predictions are still
below the current observational limits, but not by a large factor. This makes
us optimistic that both the CIB and its angular structure can be measured in
the upcoming years.  We also computed the mean flux in the background light
from this evolution in the visible bands, $B$ and $R$, and find that the
recent positive measurements of the background at these bands require high
redshift of galaxy formation, $z_F \geq 5$.

AK acknowledges support from NASA Long Term Space Astrophysics grant.

\clearpage

{\bf Figure Captions}
\noindent

{\bf Figure 1}: Spectra for various galaxy types at 14 Gyr (solid lines) and
at 1 Gyr (dotted lines). The evolution with morphological type is stronger in
the region below 4000 \AA because the main sequence  for the new
born stars lies in this region. As expected after 1 Gyr since the first burst
of star formation all morphological types have very similar spectra.

{\bf Figure 2}: The left panel shows the change with metallicity ($Z$)
in $L_B$ for an $L_*$ early
type galaxy for 3 different ages. The middle panel
shows a similar plot but for $L_K$. The right panel shows the combined
evolution of $L_B$ and $L_K$ for the metallicity range plotted in the previous
panels. The fact that $L_B \propto L_K^{0.9}$ implies that the properties of
the fundamental plane can be reproduced by systematic variations
 in metallicity with mass
for early type galaxies.

{\bf Figure 3}: The evolution with time for B,I and K magnitudes (shown
in arbitrary units) and $M/L_B$ (shown in solar units) for various galaxy types. 

{\bf Figure 4}: Evolution of the comoving luminosity density vs. time at 
2800 \AA, 4400 \AA and 1 $\mu$m for all galaxy types weighted with the CfA
fractions.  

{\bf Figure 5}: The redshift evolution of the global star formation rate in
the universe predicted for our models for late type galaxies (dashed line) and
all galaxies (solid line). The data are from \scite{Madau_97} and the square 
is from the SCUBA/HDF \scite{scuba_98}. The agreement is excellent with the 
new star formation rates derived from the SCUBA/HDF.

{\bf Figure 6}: The predicted evolution of the averaged metallicity in the
Universe for late type galaxies as predicted in our models is compared with
\pcite{Pettini+97} data for four different cosmologies.

{\bf Figure 7}: Evolution of the comoving luminosity density vs. redshift at
2800 \AA, 4400 \AA and 1 $\mu$m for all galaxy types weighted with the CfA
fractions plotted for 3 different redshifts of galaxy formation.  The data are
from \scite{Lilly+96}, \scite{Gallego+96}, \scite{Connolly+97}.  In all cases 
the
agreement between data and models is excellent, except for the case
$\Omega_0=0.2$ and $\Omega_\Lambda=0.8$. In this case the model prediction for
the UV is below the observed data since for each $z$ the galaxies are older 
than in the Einstein De Sitter Universe.

{\bf Figure 8}: Observed differential counts in several bands and our model
predictions. The fits to the counts are excellent for open models in all four
bands. The $\Omega_0=0.2$, $\Omega_\Lambda=0.8$ universe overshoots the I
counts by a small margin, but otherwise provides good fits. 
The models have been plotted for $z_f=5$, a higher value of $z_f$
would improve the fits for open universes. The prediction for the Einstein
De Sitter Universe fail to fit the data by a large margin.

{\bf Figure 9}: Data on $I-K$ vs $K$ from \pcite{Moustakas+96} 
(diamonds) are compared to our models for two cosmologies. Thick lines 
correspond to our model for  E/SO and the light lines 
for the respective Sbc model. Both models were computed for a total of
$10^{11}$ M$_{\odot}$in stars and $z_f=10$. Such high value of $z_f$ was adopted in 
order to explore if the models would predict colours redder than observed. 
The figure shows that our models do not produce  very red colours
and offer a good fit to the data. Note that E/SOs that form at 
$z>5$ will have $K \approx 23$ and $I-K \approx 2-3$ in agreement with our 
prediction from Fig. 10.

{\bf Figure 10}: The predicted redshift distribution of counts in B and K is
compared with observational data. The models are plotted with thick
solid lines for $\Omega=0.2$; dotted lines in the bottom panels correspond to
our models with $\Omega=1$. The data are taken
 from \scite{Songaila+94} (solid line histogram in the upper-left panel),
\scite{Glazebrook+95} (solid line histogram in the upper-right panel) and \scite{Cowie+96}
(dotted line in the upper-right panel), the agreement is good in all cases. The
two bottom panels show the predicted redshift distribution for the faintest
observed B and K counts. Note that galaxies in K counts should extend to 
redshifts of about 5.

{\bf Figure 11}: The rate of production of the CIB flux, $z dF/dz$ is plotted
vs $z$ for the four redshifts of galaxy formation written above each
panel. Solid lines correspond to $J$ band, dotted to $K$ and dashed to
$L$. Thin lines of each type correspond to the Einstein De Sitter Universe and
thick lines to $\Omega=0.2$ and zero cosmological constant. The first peak
correspond to the initial burst of star formation with the following dip
coming from emission redshifted from 3000-4000 \AA$\,$.

{\bf Figure 12}: The total flux for background light produced by the early
evolution of galaxies is plotted in eight bands: $B,R,I,J,H,K,L,M$. 
Solid lines
correspond to the Einstein De Sitter Universe, dashed-dotted lines to the open Universe
with $\Omega=0.2$, dashes to the flat Universe with
$(\Omega,\Omega_\Lambda)=(0.2,0.8)$ and dotted lines to
$(\Omega,\Omega_\Lambda)=(0.4,0.6)$. Thick lines correspond to $z_f=10$ and
thin lines to $z_f=3$. The plots are shown for the two values of the Hubble
constant written above each panel. The dc levels of the extragalactic diffuse
light from Bernstein (1998) are shown as filled circles with error bars.

{\bf Figure 13}: Top four panels: plot the typical order-of-magnitude
fluctuation in the CIB flux, $\sqrt{q^2P_2(q)/2\pi}$ on scale $\pi/q$. It is
plotted vs $q^{-1}$ in arcmins for the four near-IR bands shown on top of each
panel. Thin solid lines correspond to the Einstein De Sitter Universe; dotted
lines to the open Universe with $\Omega=0.2$; dashed and dashed-dotted lines
correspond to the flat Universe with $(\Omega,\Omega_\Lambda)=(0.2,0.8)$ and
(0.4,0.6) respectively. All thin lines are drawn for $H_0=50$ km/sec/Mpc. Two
lines of each type correspond to clustering being stable in proper and
comoving coordinates from bottom up. Thick solid line corresponds to the
Einstein De Sitter Universe with $H_0=80$ km/sec/Mpc.\\ Bottom four panels
show $d[q^2P_2(q)]/dz$ computed from eq.(5) with $\Psi^2(z)d(H_0t)/dz =1$. The
four panels correspond to the four near-IR bands written above the top panel.
Solid lines correspond to $q^{-1}=10$ arcmin and dotted lines to $q^{-1}=60$
arcmin.  Thick lines of each type correspond to $z_f=10$ and thin lines to
$z_f=5$.

\clearpage
\begin{figure}
\centering
\leavevmode
\epsfxsize=1.0
\columnwidth
\epsfbox{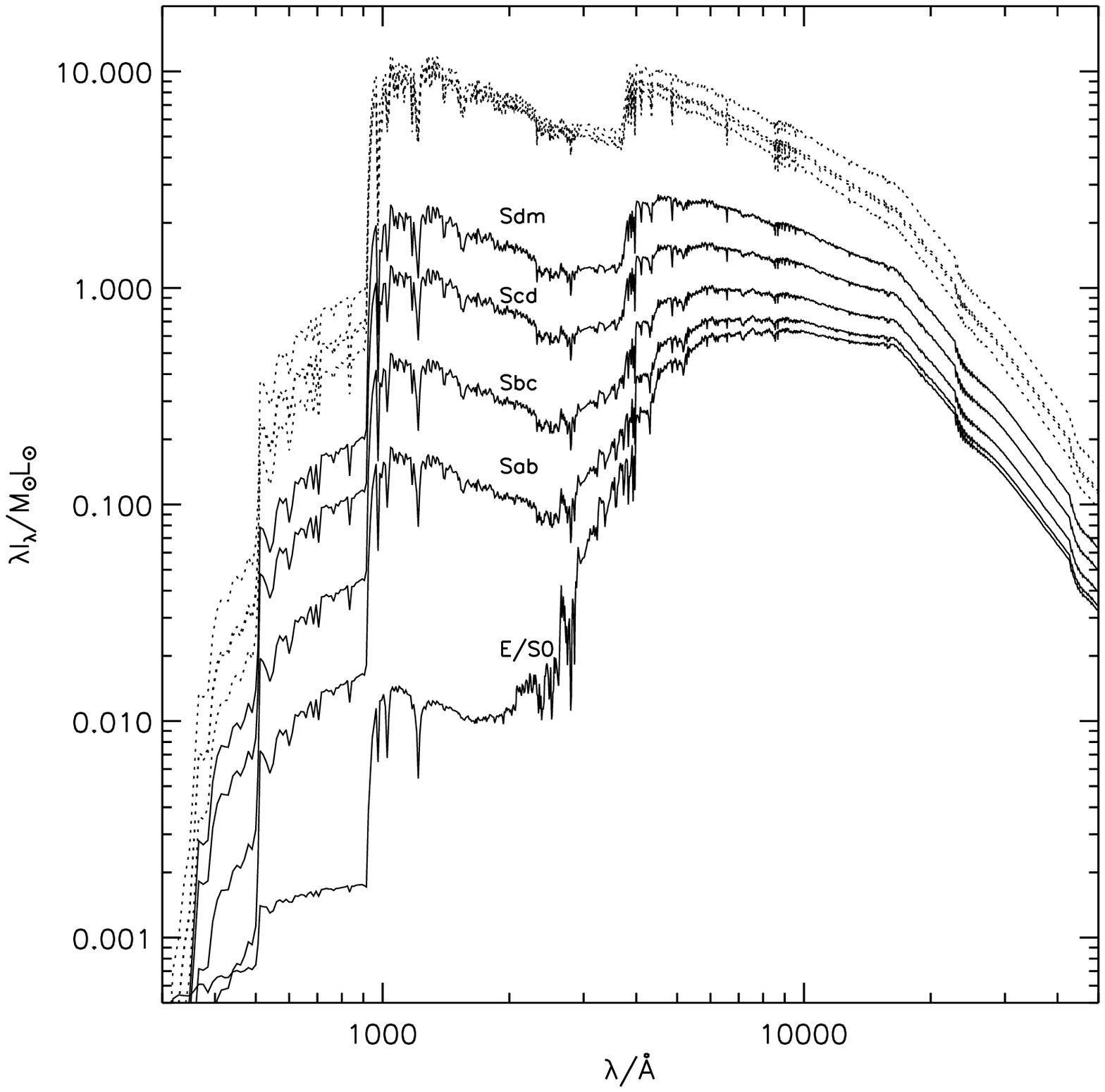}
\caption[]{}
\end{figure}

\clearpage
\begin{figure}
\centering
\leavevmode
\epsfxsize=1.0
\columnwidth
\epsfbox{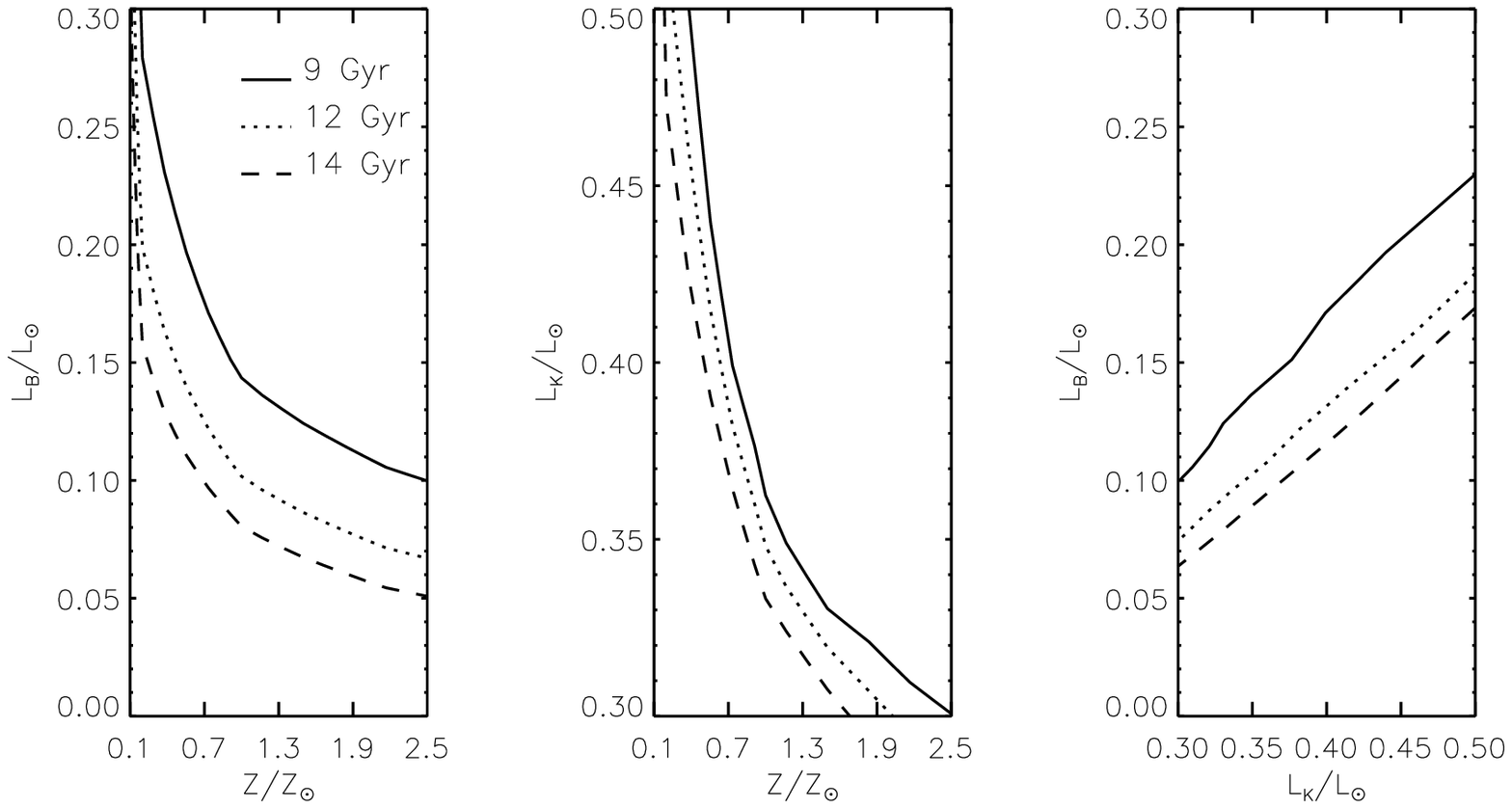}
\caption[]{}
\end{figure}

\clearpage
\begin{figure}
\centering
\leavevmode
\epsfxsize=1.0
\columnwidth
\epsfbox{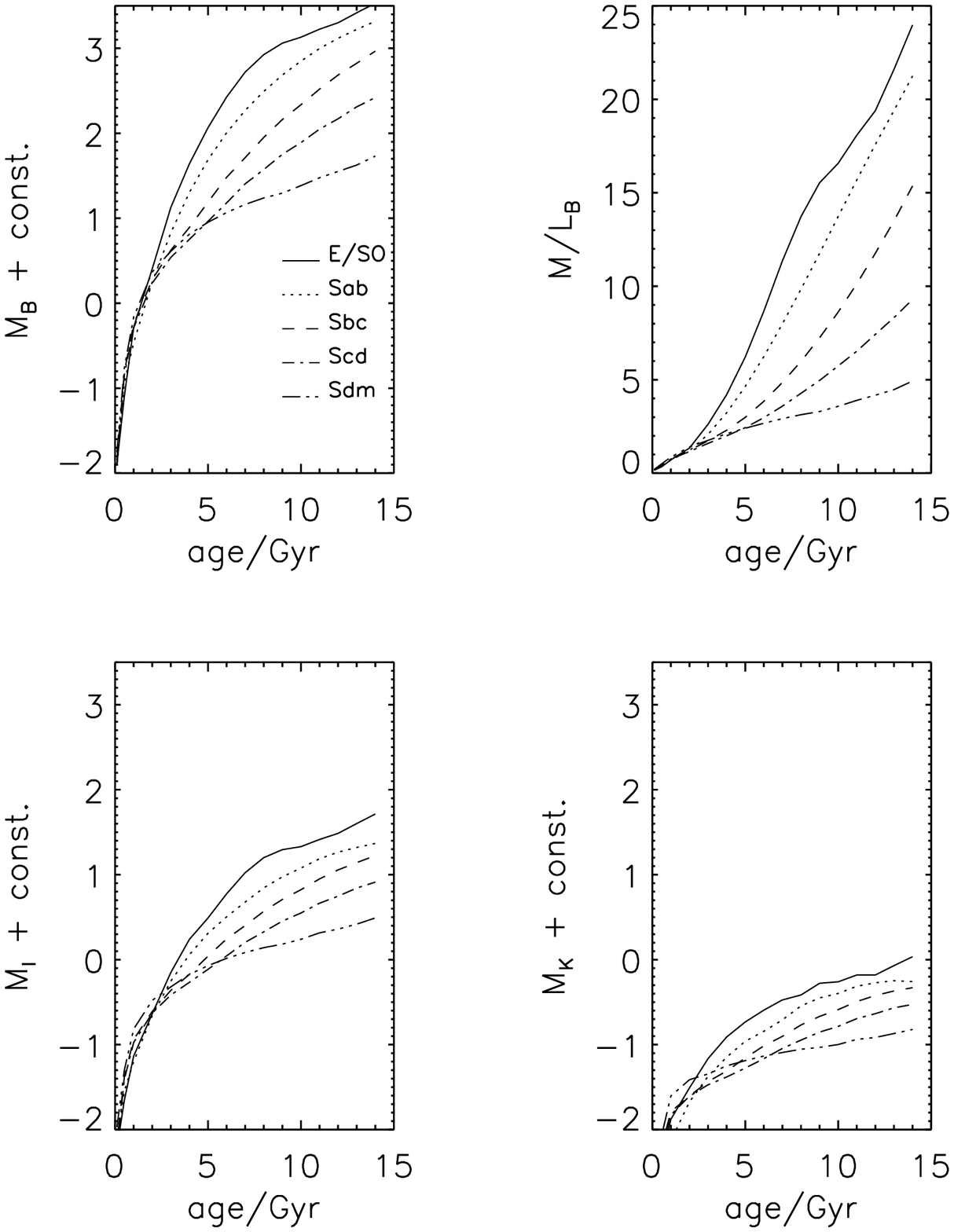}
\caption[]{}
\end{figure}

\clearpage
\begin{figure}
\centering
\leavevmode
\epsfxsize=1.0
\columnwidth
\epsfbox{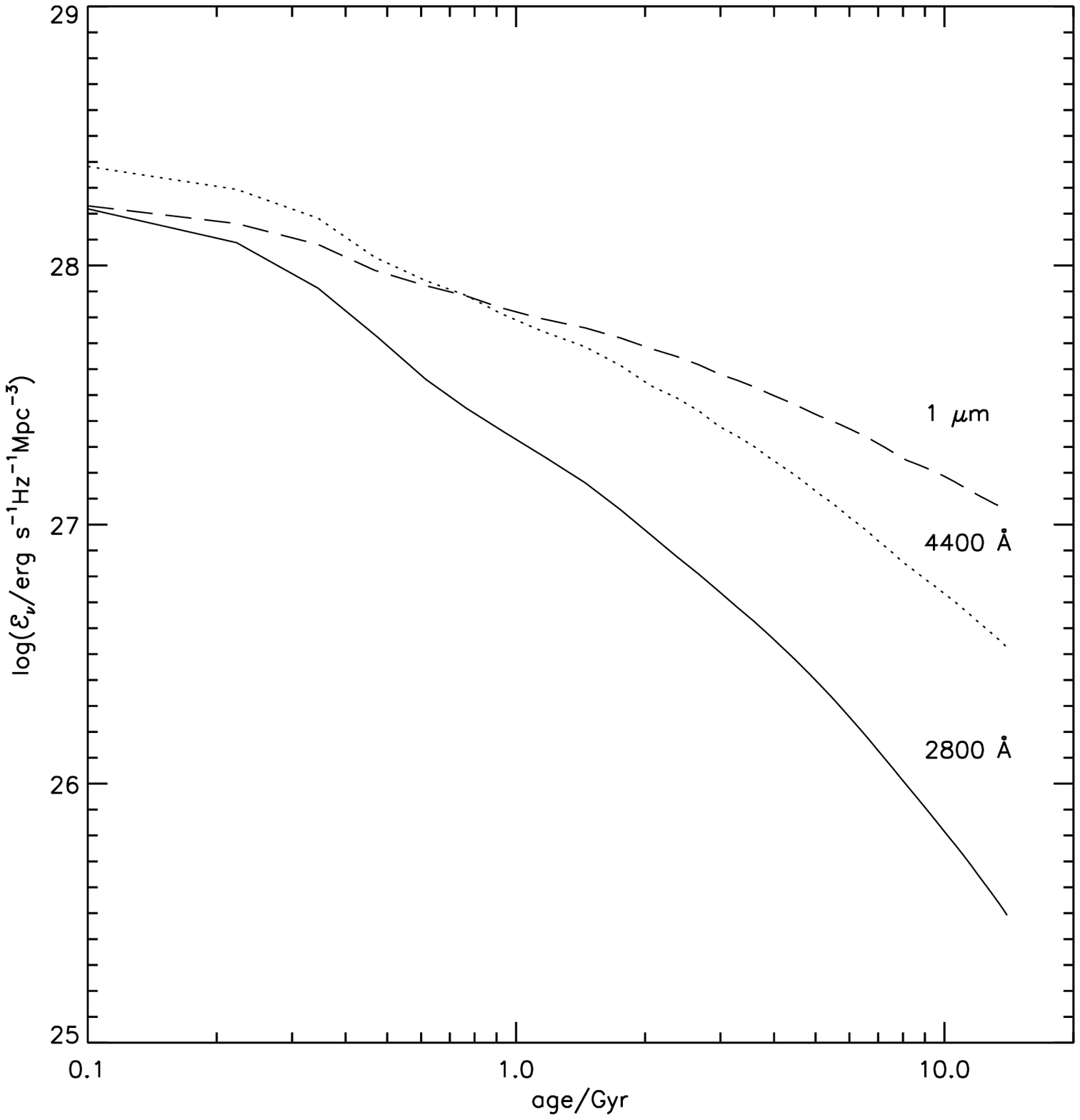}
\caption[]{}
\end{figure}

\clearpage
\begin{figure}
\centering
\leavevmode
\epsfxsize=1.0
\columnwidth
\epsfbox{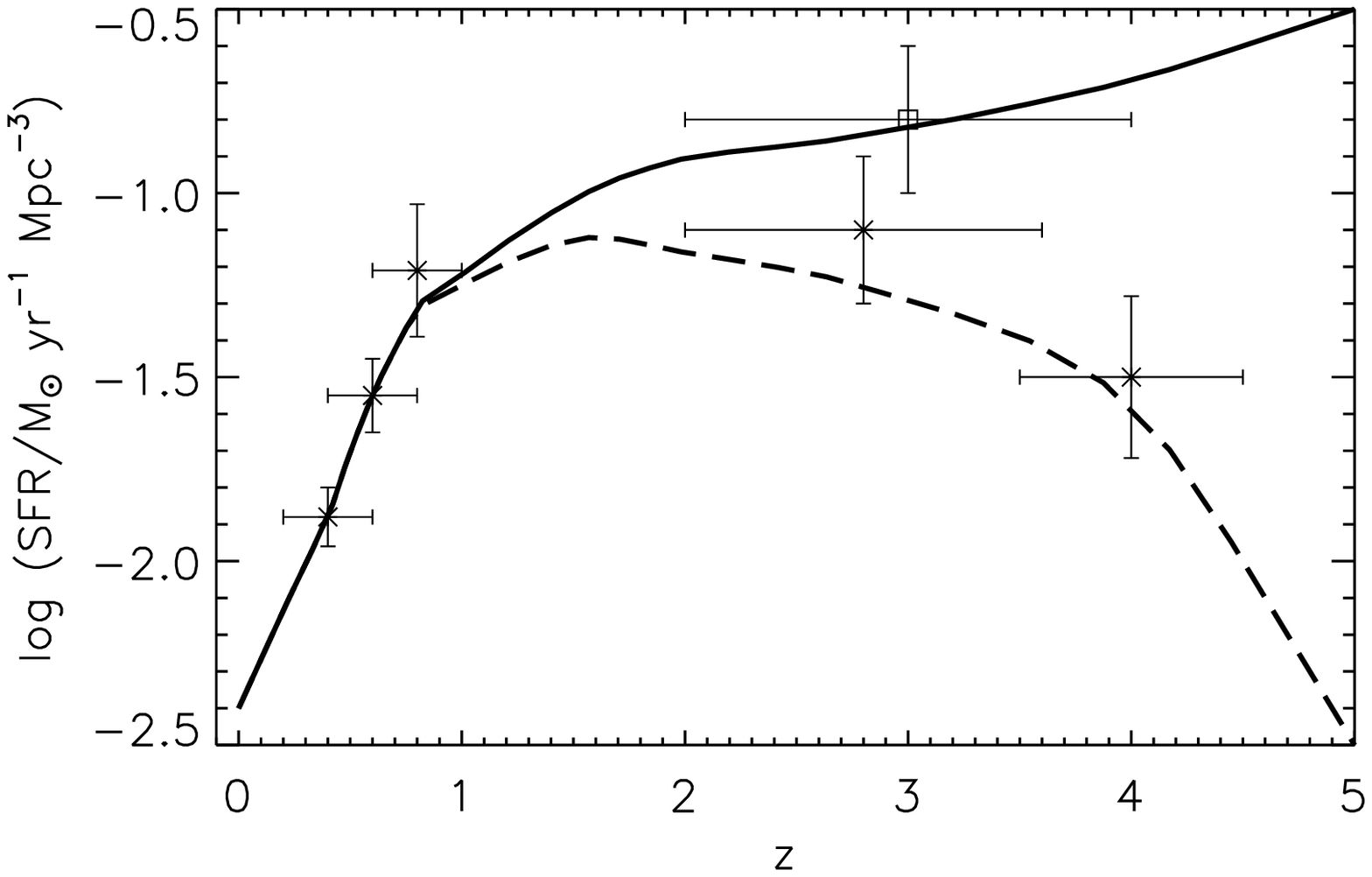}
\caption[]{}
\end{figure}

\clearpage
\begin{figure}
\centering
\leavevmode
\epsfxsize=1.0
\columnwidth
\epsfbox{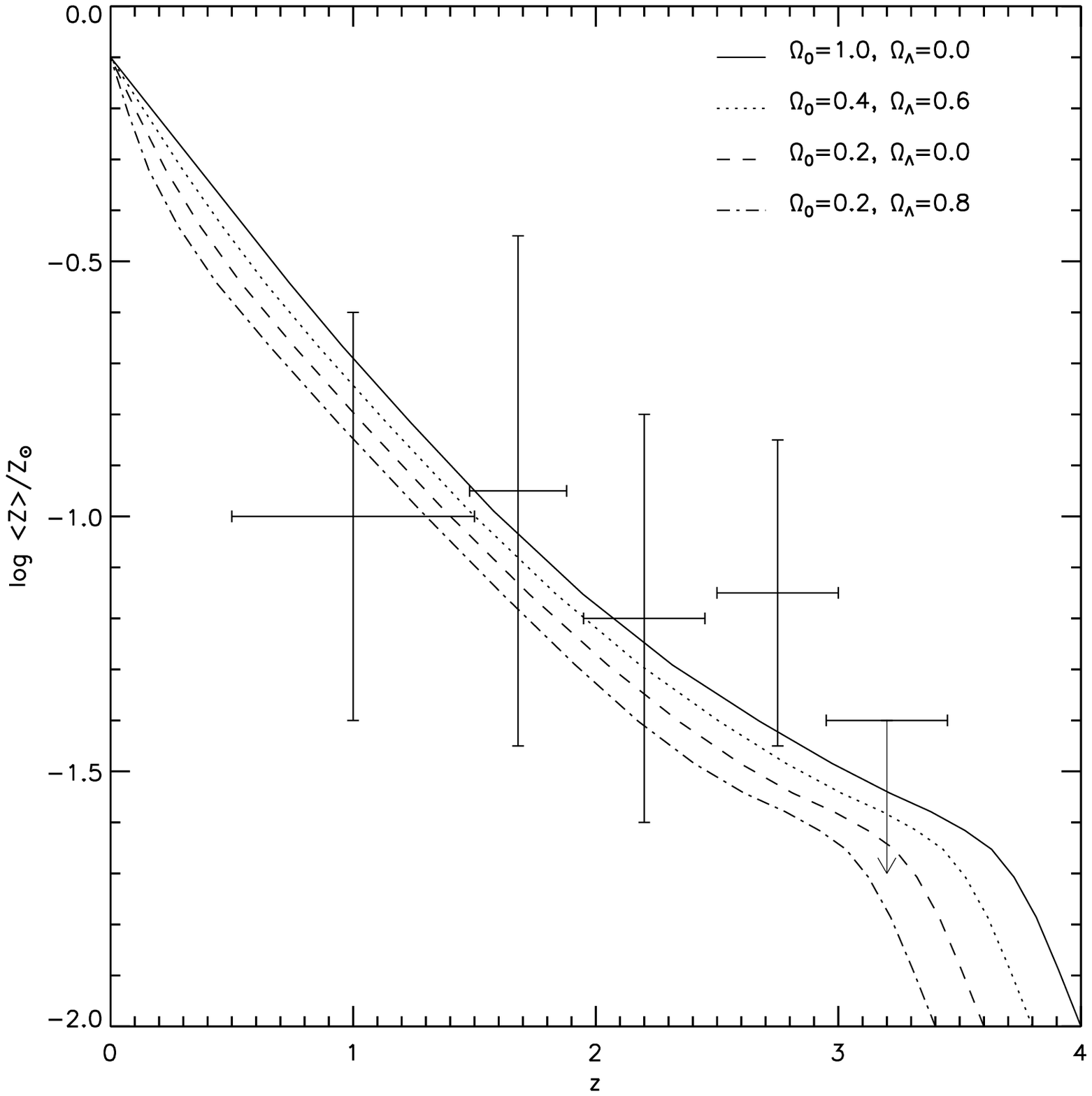}
\caption[]{}
\end{figure}

\clearpage
\begin{figure}
\centering
\leavevmode
\epsfxsize=1.0
\columnwidth
\epsfbox{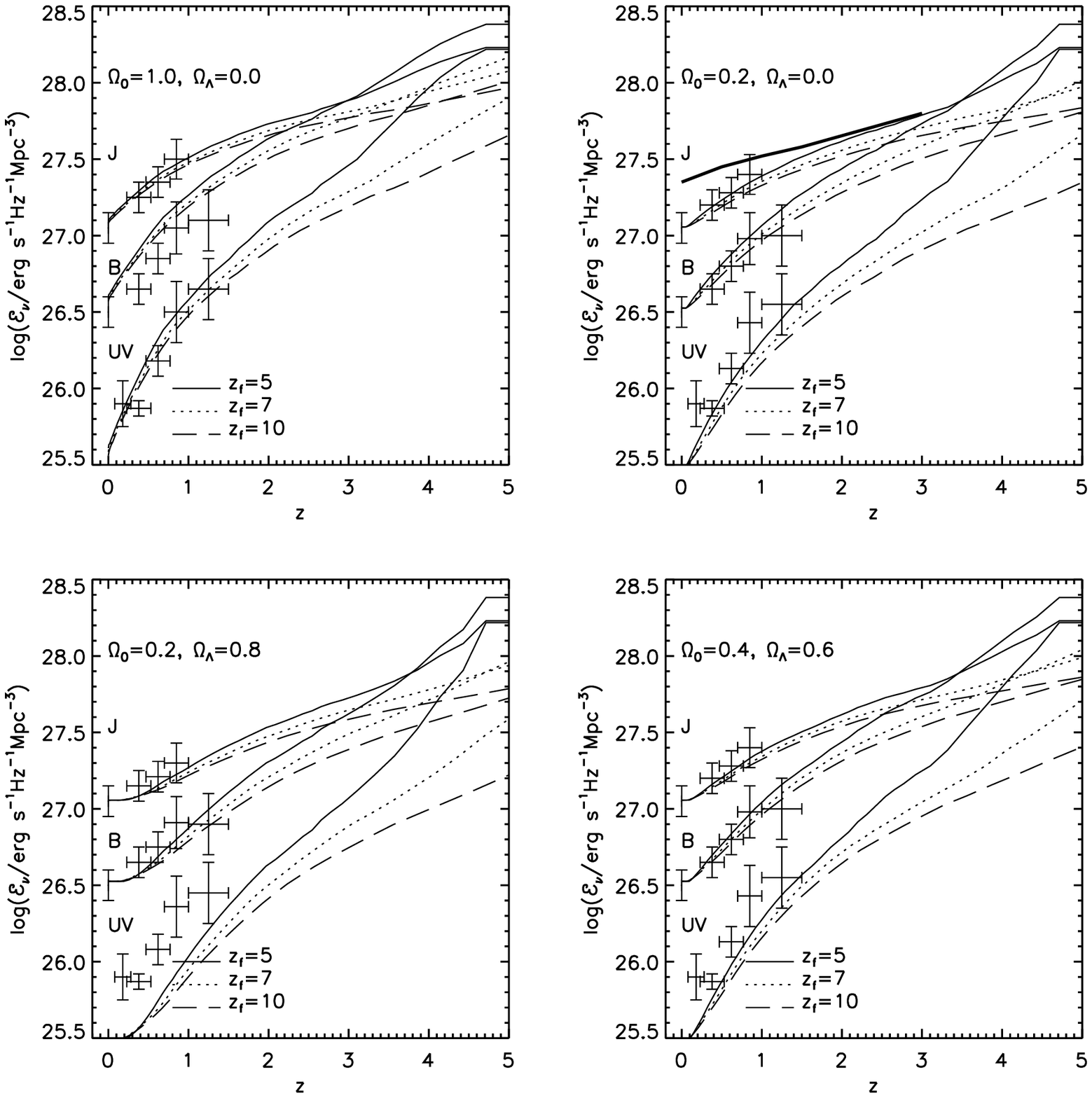}
\caption[]{}
\end{figure}

\clearpage
\begin{figure}
\centering
\leavevmode
\epsfxsize=1.0
\columnwidth
\epsfbox{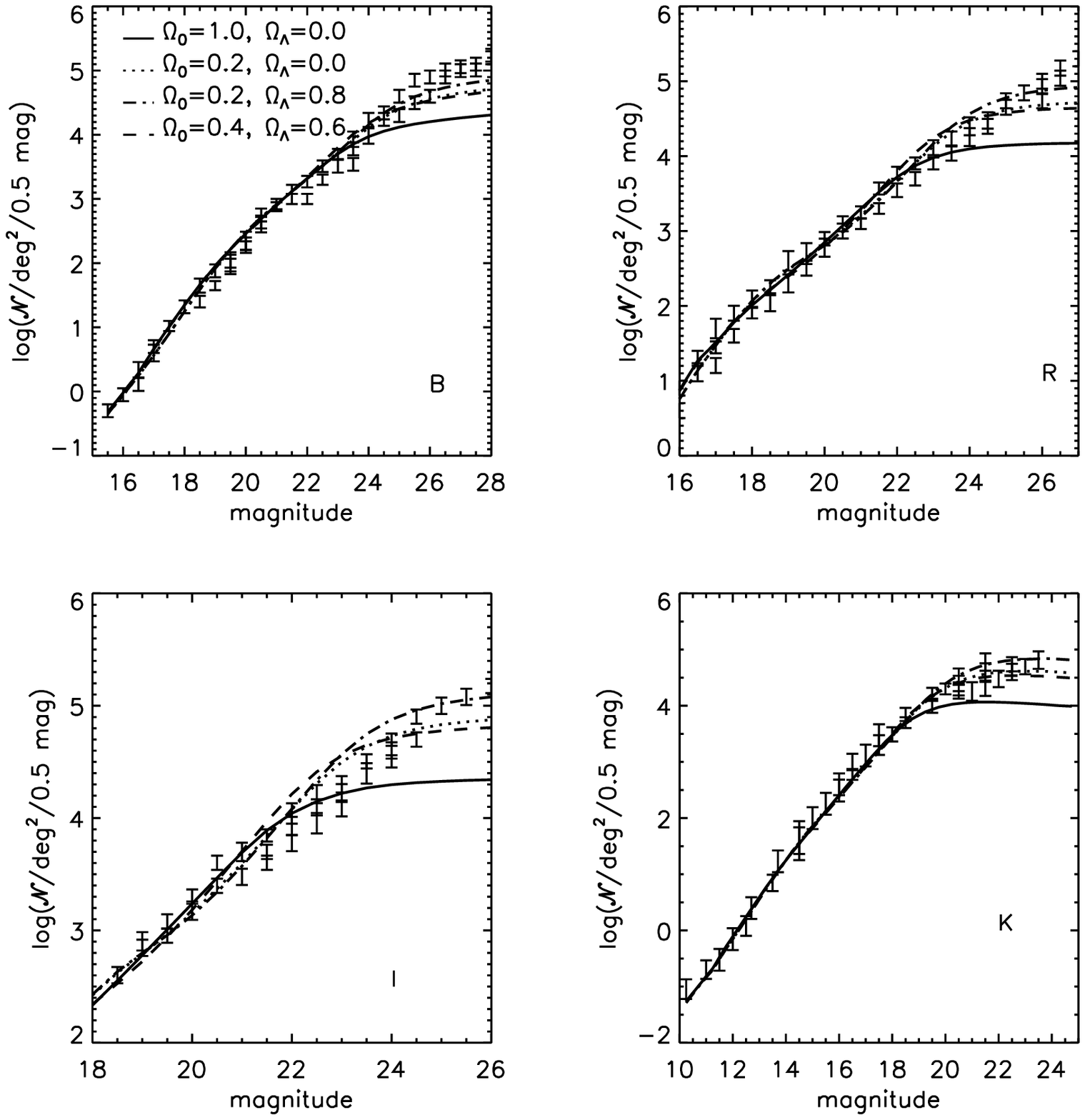}
\caption[]{}
\end{figure}
\clearpage

\clearpage
\begin{figure}
\centering
\leavevmode
\epsfxsize=1.0
\columnwidth
\epsfbox{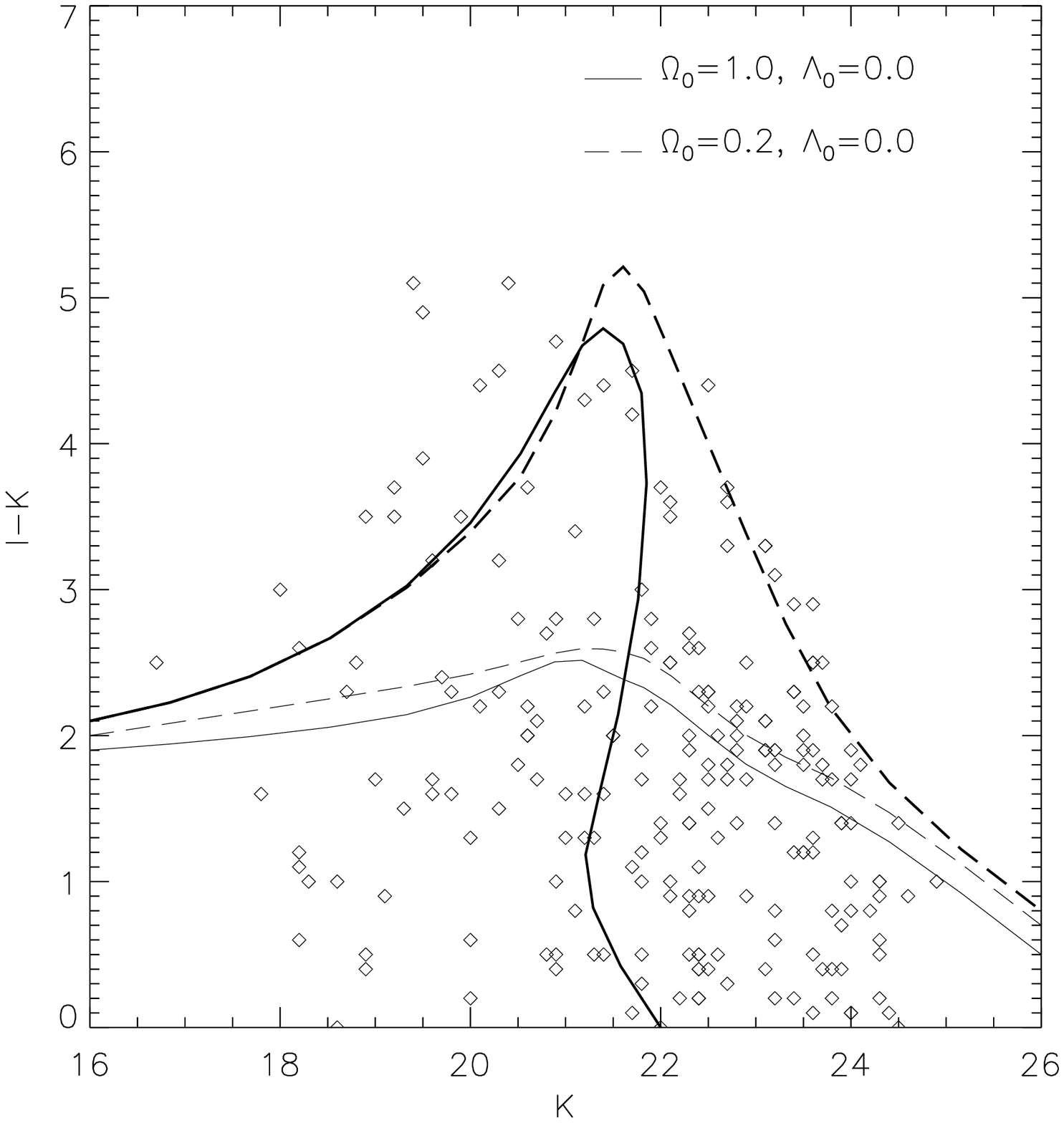}
\caption[]{}
\end{figure}
\clearpage

\begin{figure}
\centering
\leavevmode
\epsfxsize=1.0
\columnwidth
\epsfbox{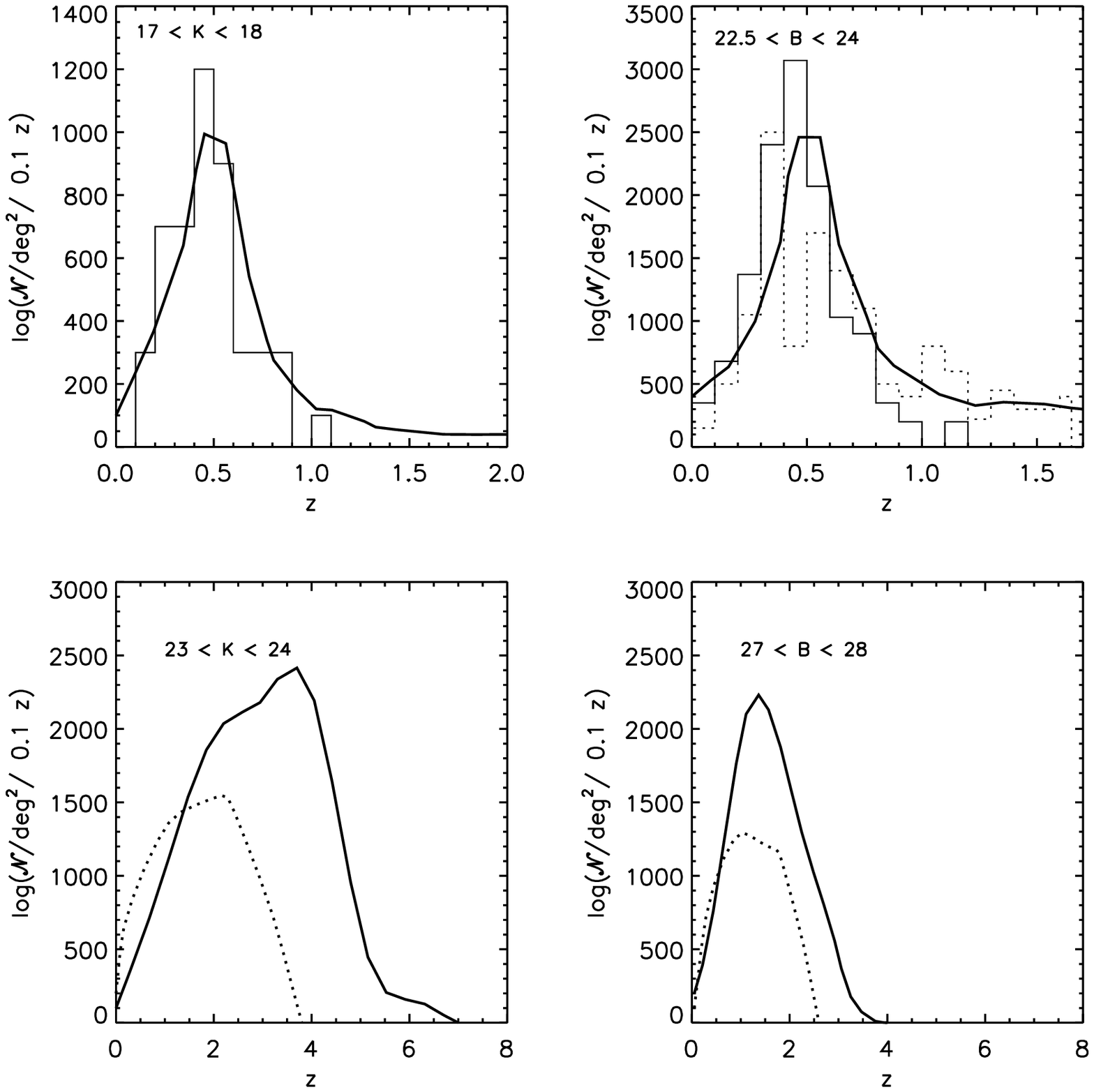}
\caption[]{}
\end{figure}

\clearpage
\begin{figure}
\centering
\leavevmode
\epsfxsize=1.0
\columnwidth
\epsfbox{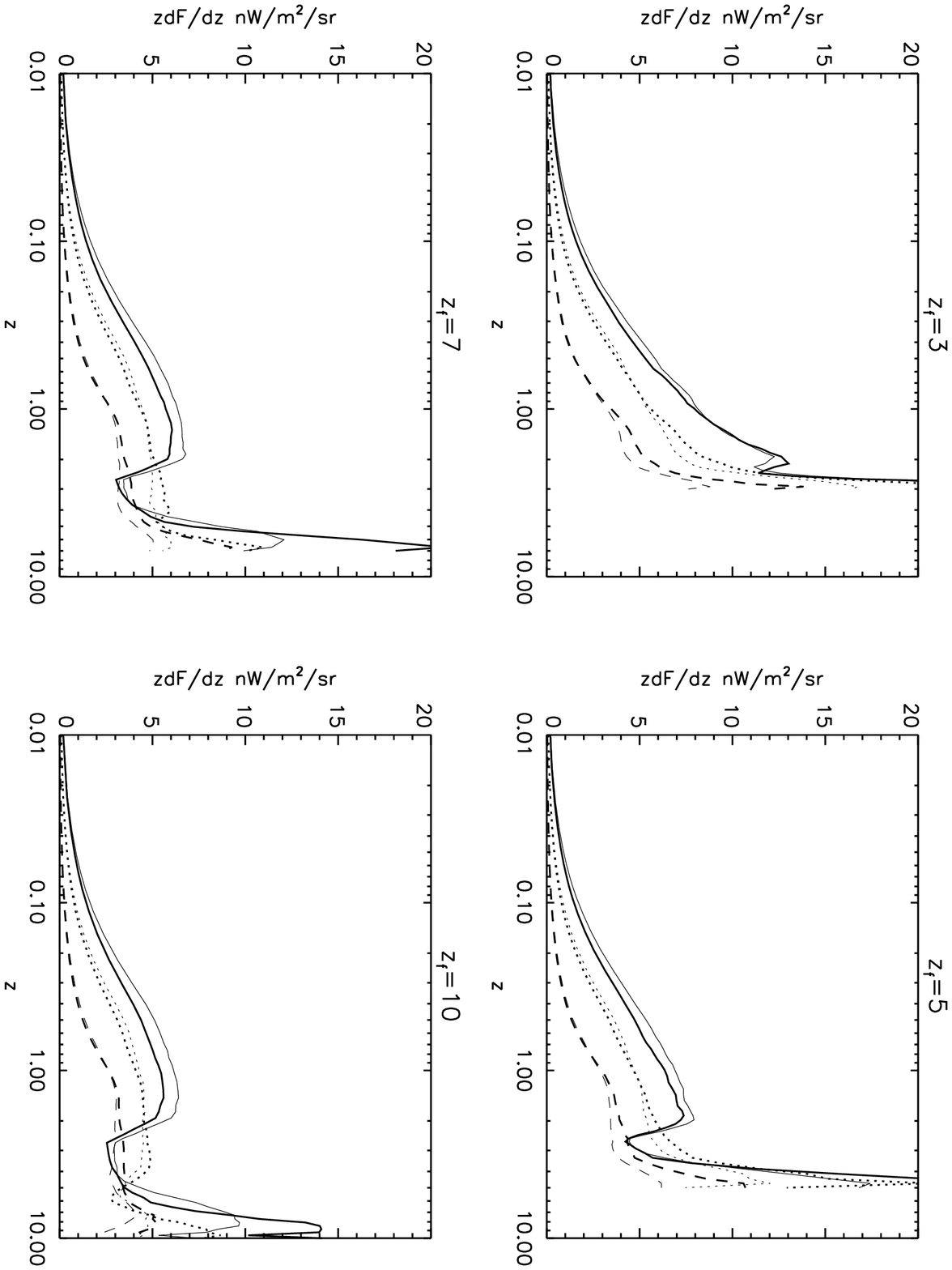}
\caption[]{}
\end{figure}

\clearpage
\begin{figure}
\centering
\leavevmode
\epsfxsize=1.0
\columnwidth
\epsfbox{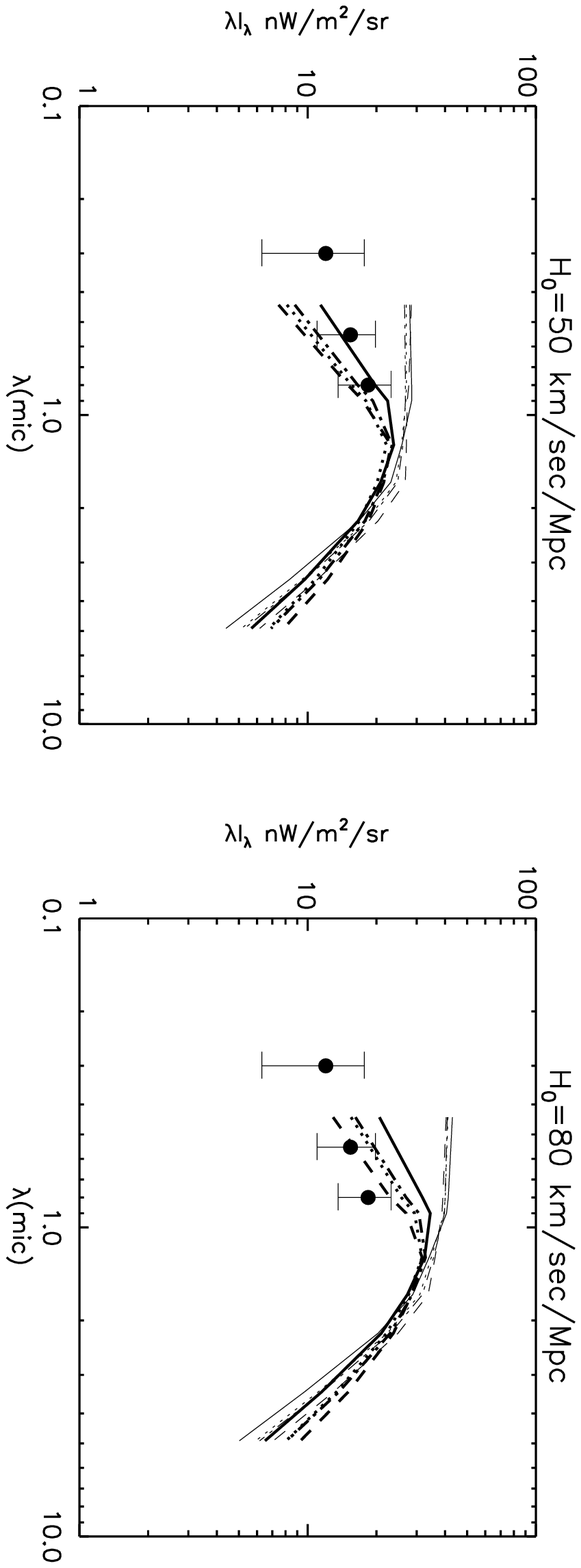}
\caption[]{}
\end{figure}

\clearpage
\begin{figure}
\centering
\leavevmode
\epsfxsize=1.0
\columnwidth
\epsfbox{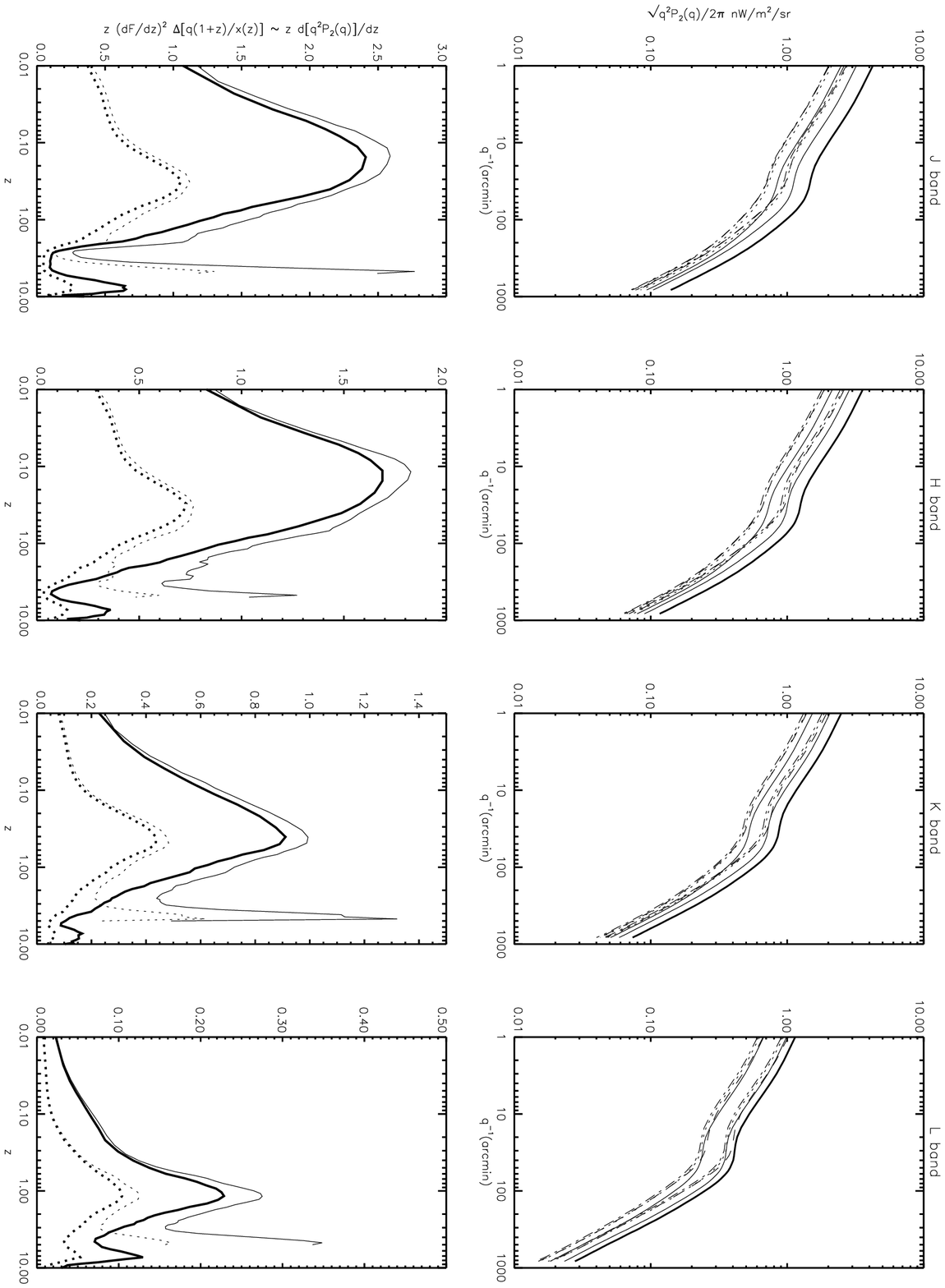}
\caption[]{}
\end{figure}

\end{document}